\pdfoutput=1

\documentclass[11pt,a4paper]{article}

\usepackage[utf8]{inputenc}
\usepackage{amsmath}
\usepackage{amsthm}
\usepackage{amssymb}
\usepackage{url}
\usepackage[english]{babel}
\usepackage{bbold}
\usepackage{slashed}
\usepackage{multirow}
\usepackage{xspace}
\usepackage{booktabs}
\usepackage{graphicx}
\usepackage{verbatim}
\usepackage{subfigure}
\usepackage{cite}
\usepackage{geometry}
\usepackage{xcolor}
\usepackage{color}
\usepackage{float}

\usepackage{siunitx}
\usepackage{hyperref}
\hypersetup{colorlinks=true,urlcolor=blue,linkcolor=red,citecolor=green!60!black}

\renewcommand{\baselinestretch}{1.2}

\newcommand\w[1]{_{\mathrm{#1}}}

\newcommand\abs[1]{\lvert#1\rvert}
\newcommand\unit[1]{~\mathrm{#1}\xspace}
\newcommand\eV{\unit{eV}}

\newcommand\GeV{\unit{GeV}}

\newcommand\PeV{\unit{PeV}}
\newcommand{\trans}{^{\mathrm T}}
\newcommand\MSbar{$\overline{\mathrm{MS}}$\xspace}
\newcommand{\yydag}{(y y^\dagger)}
\newcommand{\ph}{{\phantom{*}}}
\newcommand\ii{\mathrm{i}}
\newcommand{\pTr}[1]{\Tr\bigl(#1\bigr)}
\newcommand\NRI{N^{\mathrm R}_I}
\newcommand\mD{m\w D}
\renewcommand{\Re}{\mathop{\mathrm{Re}}}
\renewcommand{\Im}{\mathop{\mathrm{Im}}}

\DeclareMathOperator{\Tr}{\mathrm{Tr}}
\DeclareMathOperator{\Order}{\mathcal{O}}
\DeclareMathOperator{\sign}{\mathrm{sign}}

\begin{document}


\thispagestyle{empty}

\noindent~

\vskip 1.5cm

\begin{center}

{\LARGE\bf Type-I Seesaw as the Common Origin of Neutrino\\\medskip Mass, Baryon Asymmetry, and the Electroweak Scale}

\vskip 2cm

\renewcommand*{\thefootnote}{\fnsymbol{footnote}}

{\large
Vedran~Brdar,\,$^{a,\,\hspace{-0.25mm}}$\footnote{\href{mailto:vbrdar@mpi-hd.mpg.de}{vbrdar@mpi-hd.mpg.de}}
Alexander~J.~Helmboldt,\,$^{a,\,\hspace{-0.25mm}}$\footnote{\href{mailto:alexander.helmboldt@mpi-hd.mpg.de}{alexander.helmboldt@mpi-hd.mpg.de}}
Sho~Iwamoto,\,$^{b,\,c,\,\hspace{-0.25mm}}$\footnote{\href{mailto:sho.iwamoto@pd.infn.it}{sho.iwamoto@pd.infn.it}}
and Kai~Schmitz\,$^{b,\,c,\,\hspace{-0.25mm}}$\footnote{\href{mailto:kai.schmitz@pd.infn.it}{kai.schmitz@pd.infn.it}}
}\\[3mm]
{\it{
$^{a}$ Max-Planck-Institut f\"ur Kernphysik, Saupfercheckweg 1, 69117 Heidelberg, Germany\\
$^{b}$ Universit\`a degli Studi di Padova, Via Marzolo 8, 35131 Padua, Italy\\
$^{c}$ INFN, Sezione di Padova, Via Marzolo 8, 35131 Padua, Italy}}

\end{center}

\vskip 1cm

\renewcommand*{\thefootnote}{\arabic{footnote}}
\setcounter{footnote}{0}


\begin{abstract}


\noindent
The type-I seesaw represents one of the most popular extensions of the Standard Model.
Previous studies of this model have mostly focused on its ability to explain neutrino oscillations as well as on the generation of the baryon asymmetry via leptogenesis.
Recently, it has been pointed out that the type-I seesaw can also account for the origin of the electroweak scale due to heavy-neutrino threshold corrections to the Higgs potential.
In this paper, we show for the first time that all of these features of the type-I seesaw are compatible with each other.
Integrating out a set of heavy Majorana neutrinos results in small masses for the Standard Model neutrinos; baryogenesis is accomplished by resonant leptogenesis; and the Higgs mass is entirely induced by heavy-neutrino one-loop diagrams, provided that the tree-level Higgs potential satisfies scale-invariant boundary conditions in the ultraviolet.
The viable parameter space is characterized by a heavy-neutrino mass scale roughly in the range $10^{6.5\cdots7.0}\GeV$ and a mass splitting among the nearly degenerate heavy-neutrino states up to a few TeV.
Our findings have interesting implications for high-energy flavor models and low-energy neutrino observables.
We conclude that the type-I seesaw sector might be the root cause behind the masses and cosmological abundances of all known particles.
This statement might even extend to dark matter in the presence of a keV-scale sterile neutrino.


\end{abstract}


\setcounter{page}{0}

\newpage

\setcounter{page}{1}

{\hypersetup{linkcolor=black}\renewcommand{\baselinestretch}{1}\tableofcontents}


\section{Introduction}
\label{sec:intro}


\subsection{The Dirac-neutrino option}


The \textit{Standard Model} (SM) describes neutrinos in terms of massless \textit{left-handed} (LH) Weyl fermions.
The observation of neutrino flavor oscillations, however, points at nonvanishing neutrino masses, which provides direct experimental evidence for \textit{new physics} (NP) \textit{beyond the Standard Model} (BSM)~\cite{Tanabashi:2018oca}.
One straightforward way of explaining nonzero neutrino masses is to supplement the Standard Model by massless \textit{right-handed} (RH) neutrinos $N^{\rm R}_I$ that transform as complete singlets under the SM gauge group.
The presence of \textit{RH neutrinos} (RHNs) in the theory then allows one to write down a Yukawa term that couples LH and RH neutrinos to the SM Higgs doublet $\phi = \left(\phi_+,\phi_0\right)^{\rm T}$,
\begin{align}
\label{eq:LDirac}
\mathcal{L}_N^{\rm D} = \frac{\ii}{2}\,\overline{N_I^{\rm R}}\,\slashed{\partial}\,N_I^{\rm R} - y_{I\alpha}\,\overline{N_I^{\rm R}}\,\tilde\phi^\dagger L_\alpha + \textrm{h.c.} \,,\quad I = 1,2,3 \,, \quad \alpha = e,\mu,\tau \,.
\end{align}
Here, $y_{I\alpha}$ is a matrix of complex Yukawa couplings, $L_\alpha = \left(\nu_\alpha^{\rm L},\ell_\alpha^{\rm L}\right)^{\rm T}$ represents the SM LH lepton doublet of flavor $\alpha$, and $\tilde\phi = \ii\sigma_2\,\phi^* = \left(\phi_0^*,-\phi_-\right)^{\rm T}$ denotes the hypercharge-conjugated Higgs doublet.
Equation~\eqref{eq:LDirac} sets the stage for neutrino mass generation via the standard Higgs mechanism. 
Upon \textit{electroweak symmetry breaking} (EWSB), the Higgs field acquires a nonzero \textit{vacuum expectation value} (VEV), $\sqrt{2}\left<\phi_0\right> = v \simeq 246\,\textrm{GeV}$, such that LH and RH neutrinos combine into massive Dirac fermions.
This scenario is referred to as the Dirac-neutrino scenario.
In this model, the \textit{electroweak} (EW) scale $v$, which is induced by the tree-level Higgs mass parameter $\mu$, can be identified as the fundamental energy scale that determines the masses of all SM particles, \textit{i.e.}, the masses of the SM Higgs boson, EW gauge bosons, and all SM fermions.
Another attractive feature of this minimal SM extension is that it provides a possibility to explain the origin of the \textit{baryon asymmetry of the Universe} (BAU) via the so-called neutrinogenesis mechanism~\cite{Dick:1999je}.
This mechanism is based on the idea that, in the presence of RH neutrinos, the decay of heavy exotic \textit{degrees of freedom} (DOFs) in the early Universe can lead to a primordial asymmetry between LH and RH neutrinos.
The lepton number carried by the LH neutrinos, $L^{\rm L}$, is then converted into a primordial baryon number $B$ by EW sphaleron processes~\cite{Klinkhamer:1984di,Kuzmin:1985mm}.
The lepton number carried by the RH neutrinos is of equal magnitude but different sign, $L^{\rm R} = - L^{\rm L}$.
It remains sequestered from the rest of the thermal bath, until LH and RH neutrinos eventually equilibrate at late times due the Yukawa interaction in Eq.~\eqref{eq:LDirac}.


The Dirac-neutrino model manages to explain the small SM neutrino masses, offers a starting point for realistic models of baryogenesis (see, \textit{e.g.}, Ref.~\cite{Murayama:2002je}), and relates the masses of all known elementary particles to a single energy scale, \textit{i.e.}, the Higgs mass parameter $\mu$.
However, despite these achievements, it also suffers from a number of shortcomings and calls for further model building:
\begin{enumerate}
\item In order to relate tiny SM neutrino masses of $\Order(0.1)\eV$ to the Higgs VEV $v \sim 100\,\textrm{GeV}$, the RHN Yukawa couplings $y_{I\alpha}$ need to be of $\Order(10^{-12})$.
This aggravates the SM flavor puzzle. 
\item The particle content of the Dirac-neutrino model on its own is not sufficient to realize successful baryogenesis.
In order to generate a primordial chiral neutrino asymmetry, it is necessary to extend the model by new DOFs whose masses may be as large as the energy scale of gauge coupling unification, $\Lambda_{\rm GUT} \sim 10^{16}\,\textrm{GeV}$, in \textit{grand unified theories} (GUTs).
The Dirac-neutrino scenario features, in particular, no intrinsic connection between the generation of the baryon asymmetry at high energies and the phenomenology of neutrino oscillations at low energies.
\item If the Higgs VEV is regarded as a fundamental energy scale, one would naively expect that the solutions to other SM problems, such as \textit{dark matter} (DM) or the EW hierarchy problem, should also be related to new physics at or slightly above the EW scale.
However, all experimental efforts thus far have failed to directly detect new particles beyond the Standard Model. 
This challenges the notion of the Higgs VEV as a fundamental scale and might be taken as an indication that the scale of new physics may, in fact, be vastly separated from the EW scale.
\item Equation~\eqref{eq:LDirac} is not the most general Lagrangian that is compatible with the field content of the Dirac-neutrino model. Indeed, without imposing any symmetry, one is allowed to write down a Majorana mass term for the RH neutrinos, which explicitly breaks lepton number $L$.
To forbid this term, one has to impose $L$ as an exact global symmetry, or alternatively, $B-L$ as a gauge symmetry.
This represents a model-building constraint that needs to be accounted for when embedding the Dirac-neutrino model into a more comprehensive model at high energies.
\end{enumerate}


\subsection{The Majorana-neutrino option}


The shortcomings of the Dirac-neutrino model motivate the extension of Eq.~\eqref{eq:LDirac} by a Majorana mass term for the RH neutrinos, which results in the Lagrangian of the type-I seesaw model~\cite{Minkowski:1977sc,Yanagida:1979as,Yanagida:1980xy,GellMann:1980vs,Mohapatra:1979ia},
\begin{align}
\label{eq:Lseesaw}
\mathcal{L}_N^{\rm M} = \frac{\ii}{2}\,\overline{N_I^{\rm R}}\,\slashed{\partial}\,N_I^{\rm R} - y_{I\alpha}\,\overline{N_I^{\rm R}}\,\tilde\phi^\dagger L_\alpha - \frac{1}{2}\,\overline{N_I^{\rm R}}\,M_{IJ}\left(N_J^{\rm R}\right)^{\rm C} + \textrm{h.c.}
\end{align}
Here, $M_{IJ}$ is a symmetric matrix of $L$-violating Majorana masses, which are \textit{a priori} unrelated to any other SM mass scale.
The matrix $M_{IJ}$ can always be chosen to be real and diagonal, $M_{IJ} = M_I\,\delta_{IJ}$, without loss of generality.
In the model defined by Eq.~\eqref{eq:Lseesaw}, the SM neutrinos turn into Majorana fermions upon EWSB, which is why this scenario is referred to as the Majorana-neutrino scenario.
Typically, one assumes the RHN masses to be much larger than the EW scale, $M_I \gg v$. 
The SM neutrino masses then end up being suppressed not only by small RHN Yukawa couplings, but also by the large ratio of mass scales, $v/M_I \ll 1$.
In the Majorana-neutrino scenario, it is therefore no longer necessary to assume Yukawa couplings as small as $y_{I\alpha} \sim 10^{-12}$.
Another advantage of this model is that it establishes a link between baryogenesis and low-energy neutrino phenomenology.
In the type-I seesaw, the baryon asymmetry can be generated via the leptogenesis mechanism~\cite{Fukugita:1986hr}, \textit{i.e.}, via out-of-equilibrium decays of heavy Majorana neutrinos in the early Universe.
These decays generate a primordial lepton asymmetry, which is again converted to a primordial baryon asymmetry by EW sphalerons. 
For a recent series of review papers on leptogenesis, see Refs.~\cite{Dev:2017trv,Drewes:2017zyw,Dev:2017wwc,Biondini:2017rpb,Chun:2017spz}.


However, also the Majorana-neutrino scenario comes with a number of challenges and drawbacks.
One may, \textit{e.g.}, complain that the type-I seesaw model requires the introduction of new mass parameters that are unrelated to the EW scale.
One is therefore no longer able to identify a common origin of all particle masses, as it is possible in the Dirac-neutrino scenario.
Furthermore, the large hierarchy between the mass scales $M_I$ and $v$ can lead to the destabilization of the EW scale because of large radiative corrections to the Higgs mass from the RHN sector~\cite{Vissani:1997ys}.
Consider, \textit{e.g.}, standard thermal leptogenesis, which can be shown to require RHN masses as large as $M_I \gtrsim 10^9\,\textrm{GeV}$~\cite{Davidson:2002qv,Buchmuller:2002rq,Giudice:2003jh,Buchmuller:2004nz}.
In this case, the Higgs mass is necessarily fine-tuned, which may be regarded as a naturalness problem~\cite{Clarke:2015gwa}.


A possible way out of these problems is to turn the issue of radiative corrections to the Higgs mass into a virtue. 
It has recently been pointed out that the Higgs mass parameter $\mu$ in the Higgs potential can be entirely induced by RHN threshold corrections, provided that $\mu = 0$ at tree level.
This scenario is consistent with the low-energy neutrino oscillation data and has been dubbed the ``neutrino option''~\cite{Brivio:2017dfq,Brivio:2018rzm} (see Ref.~\cite{Davoudiasl:2014pya} for related earlier work).
The main premise of the neutrino option is that the classical SM Lagrangian $\mathcal{L}_{\rm SM}$ satisfies scale-invariant boundary conditions in the \textit{ultraviolet} (UV), such that $\mu = 0$ above the RHN mass threshold.
Alongside other symmetry-based approaches, such as supersymmetry or new (strongly coupled) gauge dynamics, the concept of classical scale invariance provides a well-motivated guiding principle for the construction of BSM models~\cite{Bardeen:1995kv,Hempfling:1996ht,Meissner:2006zh,Foot:2007as,Foot:2007ay,Foot:2007iy,Meissner:2007xv,Meissner:2009gs}.
In recent years, it has been applied as a tool for BSM model building in a variety of scenarios, ranging from neutrino physics over dark matter to inflation (see, \textit{e.g.}, Refs.~\cite{Holthausen:2013ota,Lindner:2014oea,Humbert:2015epa,Oda:2015gna,Kubo:2015cna,Kubo:2015joa,Haba:2015lka,Das:2015nwk,Helmboldt:2016mpi,Jinno:2016knw,Das:2016zue,Khoze:2016zfi,Kubo:2016kpb,Kubo:2017wbv,Kubo:2018kho,Marzo:2018nov,Haruna:2019zeu}).


\begin{table}
\begin{center}
\begin{tabular}{l|c|c}
\toprule
& Dirac-neutrino option
& Majorana-neutrino option \\
\midrule
Underlying symmetry& Lepton number $\rightarrow M_I = 0$& Scale invariance $\rightarrow \mu = 0$ \\
Scale behind all SM particle masses & Higgs mass parameter $\mu$& RHN Majorana masses $M_I $     \\
Anticipated scale of new physics& $\Lambda_{\rm NP} \gtrsim \mu$& $\Lambda_{\rm NP} \gtrsim M_I$ \\
SM neutrino mass generation         & Higgs mechanism           & Type-I seesaw mechanism        \\ 
Fermion type of SM neutrinos        & Dirac fermions            & Majorana fermions              \\ 
Generation of the baryon asymmetry  & Neutrinogenesis        & \textbf{Leptogenesis (this work)} \\
\bottomrule
\end{tabular}
\caption{Properties of the Dirac-neutrino and Majorana-neutrino options.
Note how the scales $M_I$ and $\mu$ exchange their roles in both scenarios due to the different underlying symmetries.
In the Dirac-neutrino scenario, the RHN masses are absent and the masses of all known particle are set by the tree-level Higgs mass parameter.
The Majorana-neutrino option is based on the reversed situation.
As a consequence, the scale of new physics is expected to be much larger in the Majorana-neutrino scenario than in the Dirac-neutrino scenario.
In this paper, we show that the generation of neutrino masses and the EW scale in the Majorana-neutrino scenario is compatible with leptogenesis.}
\label{tab:models}
\end{center}
\end{table}


It is important to note that the RHN mass terms in Eq.~\eqref{eq:Lseesaw} explicitly break scale invariance.
In the context of the neutrino option, classical scale invariance should therefore be regarded as a working assumption.
One open question is why classical scale invariance should be a good symmetry of the SM Lagrangian $\mathcal{L}_{\rm SM}$ but not of the seesaw Lagrangian $\mathcal{L}_N^{\rm M}$.
Another point is that one ultimately has to explain how classical scale invariance at low energies can emerge as the remnant of a full-fledged quantum symmetry at high energies~\cite{Tavares:2013dga}.
These aspects of the neutrino option require further investigation.
A first step in this direction has been made in Ref.~\cite{Brdar:2018vjq}, which illustrates how the RHN masses in Eq.~\eqref{eq:Lseesaw} can be generated via the spontaneous breaking of scale invariance in a theory that initially preserves conformal symmetry at the level of the entire classical Lagrangian.
Conformal symmetry breaking in this model may be associated with a first-order phase transition in the early Universe, which could give rise to an observable signal in gravitational waves~\cite{Brdar:2018num}.


\subsection{Type-I seesaw as the origin of mass and matter}


In this paper, we will stick to the original formulation of the Majorana-neutrino option in Refs.~\cite{Brivio:2017dfq,Brivio:2018rzm} and not attempt to embed it into a BSM model that is fully scale-invariant at the classical level.
In this sense, the \textit{ad hoc} assumption of scale-invariant boundary conditions in the SM sector can be seen as being on the same footing as the \textit{ad hoc} assumption of lepton number conservation in the Dirac-neutrino model.
Both the Dirac-neutrino and Majorana-neutrino models offer no intrinsic explanation for the symmetry principles that they are based on and eventually need to be extended.
However, a fascinating consequence of replacing lepton number conservation by classical scale invariance as a guiding principle in the construction of the BSM Lagrangian is that the RHN Majorana masses $M_I$ now supersede the Higgs mass parameter $\mu$ as the fundamental input scale that determines the masses of all known particles.
The Majorana-neutrino option amounts to the idea that the RHN Majorana masses first induce the EW scale, which then leads to the generation of all SM particle masses via a combination of the Higgs and type-I seesaw mechanisms.
This scenario for the origin of the SM particle masses has several advantages over the standard Higgs mechanism:
\begin{enumerate}
\item It sheds new light on the question at which energy scale one should expect to find new physics.
Provided that the EW scale is an effective scale that only comes about because of radiative corrections from the RHN sector, it is conceivable that the scale of new physics is actually to be sought at energies above the RHN mass thresholds, $\Lambda_{\rm NP} \gtrsim M_I$.
This may explain the absence of new physics in current experiments.
In this case, one could speculate that the RH neutrinos actually play the role of messenger fields that communicate with both the Standard Model and the BSM sector that is, \textit{e.g.}, responsible for the spontaneous breaking of scale invariance.
\item In the Standard Model, the Higgs mass term and its negative sign, which is crucial for EWSB, are introduced by hand.
The Majorana-neutrino option provides, by contrast, a dynamical origin for the Higgs mass term and may explain its negative sign.
The key observation is that the RHN one-loop diagrams that induce $\mu^2$ come with an overall minus sign because of the underlying Fermi-Dirac statistics. 
For an appropriate choice of the renormalization scale (see Sec.~\ref{subsec:higgs}), this negative sign eventually leads to the correct sign of the Higgs mass term.
\item In order to explain the two measured (solar and atmospheric) neutrino mass-squared differences, the seesaw sector must contain at least \textit{two RH neutrinos} (2RHNs) (see, \textit{e.g.}, Refs.~\cite{Ibarra:2003up,Guo:2006qa,Bambhaniya:2016rbb,Rink:2016vvl,Rink:2016knw}).
However, the number of RH neutrinos can easily be larger.
In particular, it may appear appealing to extend the seesaw sector by an additional RH neutrino with a mass in the keV range whose cosmological relic density accounts for the dark matter in the Universe~\cite{Dodelson:1993je,Shi:1998km,Adhikari:2016bei,Boyarsky:2018tvu,Kang:2019xuq}.
We will briefly discuss such a scenario towards the end of the paper in Sec.~\ref{subsec:discussion}.
In this case, the type-I seesaw would not only set the masses of all known particles, but also be responsible for the mass and cosmological abundance of the DM particle.
\end{enumerate}


In Refs.~\cite{Brivio:2017dfq,Brivio:2018rzm}, it has been shown that the radiative generation of the Higgs mass parameter in the type-I seesaw typically requires RHN masses of $\Order(10^7)\GeV$.
In this case, the RHN Yukawa couplings that are necessary to explain the SM neutrino masses via the type-I seesaw mechanism are too small to allow for baryogenesis via standard thermal leptogenesis.
This can also be expressed by saying that RHN masses of $\Order(10^7)\GeV$ violate the \textit{Davidson-Ibarra} (DI) bound, $M_I \gtrsim 10^9\,\textrm{GeV}$, on the RHN mass scale for standard thermal leptogenesis~\cite{Davidson:2002qv}.
For this reason, it has been argued that the neutrino option is not compatible with the simplest (vanilla) version of thermal leptogenesis.


In this paper, we are, however, going to show that the type-I seesaw does manage to simultaneously generate SM neutrino masses, the EW scale, \textit{and} the baryon asymmetry of the Universe, provided that baryogenesis proceeds via resonant leptogenesis~\cite{Pilaftsis:1997dr,Pilaftsis:1997jf,Pilaftsis:2003gt}.
In this leptogenesis scenario, the $CP$ asymmetry in RHN decays is resonantly enhanced because of a nearly degenerate RHN mass spectrum (see Ref.~\cite{Dev:2017wwc} for a review).
The additional gain in $CP$ asymmetry allows one to bypass the DI bound and lower the energy scale of leptogenesis down to values that are compatible with the neutrino option.
In our model, the small mass splitting among the RHN mass eigenstates corresponds to a second working assumption.
When embedding the type-I seesaw into a UV completion, one would have to show how this small mass splitting can be accounted for by a symmetry in the RHN sector (see, \textit{e.g.}, Refs.~\cite{Ellis:2002eh,Akhmedov:2003dg,Grossman:2003jv,Hambye:2004jf,Bjorkeroth:2016qsk}).
In passing, we also mention that an alternative possibility to lower the energy scale of leptogenesis would be to rely on a concerted interplay of flavor effects~\cite{Hambye:2003rt,Abada:2006fw,Nardi:2006fx,Abada:2006ea,Blanchet:2006be,Blanchet:2008pw,Antusch:2009gn,Moffat:2018wke,Moffat:2018smo}.
However, in this case, one would have to tune the tree-level SM neutrino masses against one-loop radiative corrections in order to realize successful leptogenesis~\cite{Moffat:2018wke,Moffat:2018smo}.
It is less clear to us how such a parametric cancellation of different terms in perturbation theory may be achieved by imposing a symmetry at the Lagrangian level. 
For this reason, we will not consider this possibility in this paper and focus on resonant leptogenesis instead.
In our analysis, resonant leptogenesis therefore acts as the counterpart of neutrinogenesis in the Dirac-neutrino scenario (see Tab.~\ref{tab:models} for a comparison between the Dirac-neutrino and Majorana-neutrino options).


A remarkable outcome of our analysis is the realization that the type-I seesaw may not only be responsible for the masses of all SM particles, but also for the asymmetry between matter and antimatter in the Universe.
In the context of the Majorana-neutrino option, it is therefore possible to identify the type-I seesaw as the principle cause behind the masses and cosmological abundances of all known particles.
Moreover, if the seesaw sector also contains an additional keV-scale RH neutrino, this statement can be even extended to include dark matter.
In this case, the type-I seesaw would be the origin of the masses and cosmological abundances of visible and dark matter.


The remainder of this paper is organized as follows:
In the next section, we will review the type-I seesaw and discuss how it manages to simultaneously generate SM neutrino masses (Sec.~\ref{subsec:neutrinos}), the baryon asymmetry of the Universe (Sec.~\ref{subsec:bau}), and the EW scale (Sec.~\ref{subsec:higgs}).
In Sec.~\ref{sec:results}, we will then turn to the bulk of our analysis and show how the requirements of (i) successful baryogenesis and (ii) the neutrino option allow one to constrain the parameter space of the type-I seesaw model.
We will present some analytical estimates (Sec.~\ref{subsec:estimates}), the results of a comprehensive numerical parameter scan (Sec.~\ref{subsec:scan}), and discuss the implications of our analysis for high-energy flavor models and keV-scale sterile-neutrino dark matter (Sec.~\ref{subsec:discussion}).
Section~\ref{sec:conclusions} contains our conclusions and a brief outlook.


\section{Type-I seesaw}
\label{sec:seesaw}


\subsection{Neutrino masses}
\label{subsec:neutrinos}


The type-I seesaw  Lagrangian is given in Eq.~\eqref{eq:Lseesaw}.
In this paper, we shall restrict ourselves to the minimal type-I seesaw involving only two RH neutrinos, $N_I^{\rm R}$ ($I=1,2$), for simplicity.
This assumption is consistent with the present low-energy data on neutrino oscillations as well as with the generation of the baryon asymmetry via leptogenesis.
In the 2RHN seesaw model, the Yukawa matrix $y_{I\alpha}$ in Eq.~\eqref{eq:Lseesaw} is (in general) a rank-2 matrix.
This is sufficient to explain the two known nonzero mass-squared differences in the SM neutrino sector.
At the same time, one of the three SM neutrino masses, $m_i$ ($i=1,2,3$), is always bound to vanish, $\min\left\{m_1,m_2,m_3\right\} = 0$.
Given the fact that the absolute neutrino mass scale, $m_{\rm tot} = \sum_i m_i$, has not yet been measured, this is a perfectly viable possibility at present.
Similarly, resonant leptogenesis only demands two nearly degenerate RHN mass eigenstates; the presence of a third RH neutrino is not necessarily required.%
\footnote{Baryogenesis via heavy-particle decay always requires at least two particles that contribute to loop amplitudes~\cite{Kolb:1990vq}.}


Working with only two RH neutrinos has two important advantages.
First of all, from a physical point of view, it is a relevant observation that already the minimal 2RHN seesaw succeeds in explaining neutrino masses, baryon asymmetry, and the EW scale.
It is actually not necessary to consider the standard scenario involving \textit{three RH neutrinos} (3RHNs).
This leaves room for adding a third, keV-scale RH neutrino $N_3^{\rm R}$ that would not affect the low-energy neutrino observables, but which could act as a DM candidate (see Sec.~\ref{subsec:discussion}).
Such a scenario represents a highly attractive possibility that deserves further scrutiny in future work.
A second advantage of working with only two RH neutrinos is that it leads to simplifications at the technical level.
The 3RHN seesaw model features 18 free parameters in the high-energy Lagrangian (3 RHN masses plus 9 complex Yukawa couplings minus 3 unphysical charged-lepton phases), whereas the 2RHN seesaw model only contains 11 free parameters at high energies (2 RHN masses plus 6 complex Yukawa couplings minus 3 unphysical charged-lepton phases).
At the same time, the 3RHN seesaw gives rise to 9 observable quantities at low energies (3 nonzero SM neutrino masses, 3 mixing angles, and 3 physical $CP$-violating phases), while the 2RHN seesaw only leads to 7 observables (2 nonzero SM neutrino masses, 3 mixing angles, and 2 physical $CP$-violating phases).
Fixing the masses in the RHN spectrum for the purposes of leptogenesis, this means that there is a mismatch of 6 real DOFs between high-energy and low-energy quantities in the 3RHN seesaw, but only a mismatch of 2 real DOFs in the 2RHN seesaw.
In other words, the unconstrained theory space of possible flavor models has six real dimensions in the 3RHN seesaw, while it is only two-dimensional in the 2RHN seesaw.
As a consequence, it is easier to scan over all possible flavor models in the 2RHN seesaw than in the 3RHN seesaw. 


Let us now review the mechanism of SM neutrino mass generation in the type-I seesaw model.
In the course of EWSB, the SM Higgs doublet develops a nonvanishing VEV, $\sqrt{2}\left<\phi_0\right> = v \simeq 246\,\textrm{GeV}$, which generates a matrix of complex Dirac masses, $(\mD)_{I\alpha}$, for the LH and RH neutrinos in Eq.~\eqref{eq:Lseesaw},
\begin{equation}
\mathcal{L}_N^{\rm M} \:\:\overset{\rm EWSB}{\longrightarrow}\:\:
\frac{\ii}{2}\,\overline{\NRI}\,\slashed{\partial}\,\NRI
- \left[\left(\mD\right)_{I\alpha} + y_{I\alpha}\,\phi_0\right] \overline{N_I^{\rm R}}\,\nu_\alpha^{\rm L}
+ y_{I\alpha}\,\phi_+\,\overline{N_I^{\rm R}}\,\ell_\alpha^{\rm L} - \frac{1}{2}\,\overline{N_I^{\rm R}}\,M_{IJ}\,(N_J^{\rm R})^{\rm C} + \textrm{h.c.}
\end{equation}
Here, $\phi_0$ contains the real SM Higgs boson with a mass of $m_{h_0} \simeq 125\,\textrm{GeV}$ after EWSB. The Dirac mass matrix is directly proportional to the RHN Yukawa matrix, $(\mD)_{I\alpha} = y_{I\alpha}\,v/\sqrt{2}$.
After EWSB, the Dirac and Majorana mass terms in the neutrino Lagrangian can be organized as follows,
\begin{equation}
\label{eq:M}
\mathcal{L}_N^{\rm M} \supset - \frac{1}{2}\begin{pmatrix}\overline{(v_\alpha^{\rm L})^{\rm C}} & \overline{N_I^{\rm R}}\end{pmatrix} \begin{pmatrix} \mathbb{0}_{\alpha\beta} & \left(\mD\trans\right)_{\alpha J} \\ \big(\mD\big)_{I\beta} & M_{IJ}\end{pmatrix} \begin{pmatrix}v_\beta^{\rm L} \\ (N_J^{\rm R})^{\rm C}\end{pmatrix} + \textrm{h.c.}
\end{equation}
The total neutrino mass matrix therefore corresponds to a complex symmetric $5\times5$ matrix $\mathcal{M}$.
Thus, there is a unitary matrix that diagonalizes $\mathcal{M}$ by means of a Autonne-Takagi factorization,
\begin{align}
\mathcal{M}_{\left(\alpha,I\right)\left(\beta,J\right)} = \begin{pmatrix} \mathbb{0}_{\alpha\beta} & \left(\mD\trans\right)_{\alpha J} \\ \big(\mD\big)_{I\beta} & M_{IJ}\end{pmatrix}  \:\:\longrightarrow\:\: \mathcal{D}_{(i,I)(j,J)} = \begin{pmatrix}D_{ij}^\nu & \mathbb{0}_{iJ} \\ \mathbb{0}_{Ij} & D_{IJ}^N\end{pmatrix} \,,
\end{align}
where $D^\nu$ and $D^N$ contain three light and two heavy Majorana mass eigenvalues, respectively,
\begin{align}
D_{ij}^\nu = m_i\,\delta_{ij} \,, \quad D_{IJ}^N = M_I'\,\delta_{IJ} \,.
\end{align}
Regarding the heavy neutrinos, we are able to define $0 < M_1' \leq M_2'$, without loss of generality.
However, as for the light neutrinos, we need to distinguish between the case of a \textit{normal hierarchy} (NH), $0 = m_1 < m_2 < m_3$, and the case of an \textit{inverted hierarchy} (IH), $0 = m_3 < m_1 < m_2$.
The ordering among the light neutrino mass eigenstates then determines the sign of the largest possible mass-squared difference in the light-neutrino mass spectrum, $\Delta m_{3l}^2$.
For NH, we have $\Delta m_{3l}^2 = \Delta m_{31}^2 = m_3^2 - m_1^2 = m_3^2 > 0$, while for IH, we have $\Delta m_{3l}^2 = \Delta m_{32}^2 = m_3^2 - m_2^2 = -m_2^2 < 0$.


Up to now, our discussion has been completely general.
The above results are also valid if there should be no large hierarchy among the mass eigenvalues $m_i$ and $M_I'$.
However, in the following, we will restrict ourselves to the seesaw limit, in which the RHN masses in Eq.~\eqref{eq:Lseesaw} are considerably larger than the EW scale.
In this case, the RH neutrinos decouple at high energies, such that there is no appreciable mixing among the active and sterile neutrino states at low energies.
In addition, we will also assume the RHN Majorana mass matrix to be diagonal from the outset.
In the seesaw limit, we are able to perform an approximate block diagonalization of the total mass matrix $\mathcal{M}$,
\begin{align}
\label{eq:Mblock}
\mathcal{M}_{(\alpha,I)(\beta,J)} \:\: \rightarrow \:\: \mathcal{M}_{(\alpha,I)(\beta,J)}^{\rm block} \approx \begin{pmatrix} m_{\alpha\beta} & \mathbb{0}_{\alpha J} \\ \mathbb{0}_{I\beta} & M_{IJ}\end{pmatrix} \,,
\end{align}
where we carried out a perturbative expansion in ratios of the form $(\mD)_{I\alpha}/M_{JK}$ and only kept the leading terms.
The mass matrix for the heavy Majorana neutrinos now coincides with the RHN mass matrix in Eq.~\eqref{eq:Lseesaw}.
This means that, in the seesaw limit, the heavy neutrino mass eigenvalues coincide with the RHN input masses at the Lagrangian level, $M_{IJ} = M_I\,\delta_{IJ} = D_{IJ}^N = M_I'\,\delta_{IJ}$. 
Meanwhile, we obtain the following expression for the Majorana mass matrix of the light SM neutrinos,
\begin{align}
\label{eq:mab}
m_{\alpha\beta} = - (\mD\trans)_{\alpha I}^{\vphantom{-1}}\,M_{IJ}^{-1}\,(\mD)_{J\beta}^{\vphantom{-1}} \,,
\end{align}
which reflects the fact that, in the type-I seesaw model, the masses of the light SM neutrinos are suppressed by the combination of small Yukawa couplings and large RHN masses.


The Majorana mass matrix $m_{\alpha\beta}$ in Eq.~\eqref{eq:mab} is again a complex symmetric matrix.
Thus, there is again a unitary matrix $U$ that diagonalizes $m_{\alpha\beta}$ by means of a Autonne-Takagi factorization,
\begin{align}
\label{eq:mTakagi}
(U\trans)_{i \alpha}\, m_{\alpha \beta}\, U_{\beta j} = D_{ij}^\nu = m_i\,\delta_{ij} \,.
\end{align}
The matrix $U$ relates the SM neutrino flavor eigenstates $\nu_\alpha^{\rm L}$ to the SM neutrino mass eigenstates $\nu_i$,
\begin{align}
\nu_\alpha^{\rm L} = U_{\alpha i}\,\nu_i \,, \quad \nu_i = (U^\dagger)_{i\alpha}^{\vphantom{*}}\,\nu_\alpha^{\rm L} = U_{\alpha i}^*\,\nu_\alpha^{\rm L} \,.
\end{align}
We assume a diagonal mass matrix for the charged-lepton flavors $e$, $\mu$, and $\tau$. 
This is possible because one is always able to perform unitary flavor transformations on the LH lepton doublet $L_\alpha$ as well as on the RH charged-lepton singlet $\ell_\alpha^{\rm R}$ prior to EWSB.
In this case, there will be no contributions to lepton mixing from the charged-lepton mass matrix, and the unitary matrix $U$ can be identified with the \textit{Pontecorvo--Maki--Nakagawa--Sakata} (PMNS) lepton mixing matrix~\cite{Pontecorvo:1957qd,Maki:1962mu}, $U_{\rm PMNS} = U$.
In the 2RHN seesaw model, the PMNS matrix can be parametrized as follows,
\begin{align}
U = \begin{pmatrix}
c_{12}\,c_{13} &
s_{12}\,c_{13} &
s_{13}\,e^{-i\delta} \\
-s_{12}\,c_{23} - c_{12}\,s_{13}\,s_{23}\,e^{i\delta} &
\phantom{-}c_{12}\,c_{23} - s_{12}\,s_{13}\,s_{23}\,e^{i\delta} &
s_{23}\,c_{13} \\
\phantom{-} s_{12}\,s_{23} - c_{12}\,s_{13}\,c_{23}\,e^{i\delta} & 
-c_{12}\,s_{23} - s_{12}\,s_{13}\,c_{23}\,e^{i\delta} & 
c_{23}\,c_{13}
\end{pmatrix}
\begin{pmatrix}
1 & 0 & 0 \\
0 & e^{i\sigma} & 0 \\
0 & 0 & 1
\end{pmatrix} \,.
\end{align}
Here, $s_{ij}$ and $c_{ij}$ are shorthand notations for $\sin\theta_{ij}$ and $\cos\theta_{ij}$, respectively.
$\theta_{12},\theta_{23},\theta_{12} \in\left[0,\pi/2\right)$ denote the three PMNS mixing angles, $\delta \in \left[0,2\pi\right)$ is the $CP$-violating Dirac phase, and $\sigma \in \left[0,\pi\right)$ represents the only physical $CP$-violating Majorana phase in the 2RHN seesaw.
The second Majorana phase that is present in the 3RHN seesaw, $\tau$, can always be rotated away by a phase transformation of the massless SM neutrino mass eigenstate.
In Tab.~\ref{tab:data}, we list the current experimental constraints on the neutrino observables that are accessible in experiments according to the global-fit analysis in Refs.~\cite{Esteban:2018azc,nufit}. 
Note that the $CP$-violating Majorana phase $\sigma$ is at present unconstrained.


\begin{table}
\begin{center}
\begin{tabular}{l|S[table-format=1.3]c|S[table-format=1.3]c}
\toprule
  & \multicolumn{2}{c|}{Normal hierarchy}   & \multicolumn{2}{c}{Inverted hierarchy}   \\
  & {\quad Best-fit value \;\;} & $3\,\sigma$ range
  & {\quad Best-fit value \;\;} & $3\,\sigma$ range \\
\midrule
$\,\left.\Delta m_{21}^2\right.\negthinspace/10^{-5}\eV^2$ & 7.39  & 6.79\,--\,8.01  & 7.39  & 6.79\,--\,8.01 \\
$\abs{\Delta m_{3l}^2}/10^{-3}\eV^2$                       & 2.525 & 2.431\,--\,2.622 & 2.512 & 2.413\,--\,2.606 \\
$\sin^2\theta_{12}$ & 0.310   &   0.275\,--\,0.350   & 0.310   &   0.275\,--\,0.350   \\
$\sin^2\theta_{23}$ & 0.582   &   0.428\,--\,0.624   & 0.582   &   0.433\,--\,0.623   \\
$\sin^2\theta_{13}$ & 0.02240 & 0.02044\,--\,0.02437 & 0.02263 & 0.02067\,--\,0.02461 \\
$\delta/\text{rad}$ & 3.79    &    2.36\,--\,6.39    & 4.89    &    3.42\,--\,6.13    \\
$\sigma/\text{rad}$ & {---}     &       0\,--\,$\pi$   & {---}     &       0\,--\,$\pi$   \\
\bottomrule
\end{tabular}
\caption{Best-fit values and $3\,\sigma$ confidence intervals for the low-energy neutrino observables according to the NuFIT 4.0 global-fit analysis~\cite{Esteban:2018azc,nufit}, including data on atmospheric neutrinos from Super-Kamiokande~\cite{Abe:2017aap}.
The largest mass-squared difference in the SM neutrino mass spectrum, $\Delta m^2_{3l}$, is defined as $m_3^2-m_1^2 = m_3^2 > 0$ for NH and as $m_3^2-m_2^2 = -m_2^2 < 0$ for IH.}
\label{tab:data}
\end{center}
\end{table}


The identities in Eqs.~\eqref{eq:mab} and \eqref{eq:mTakagi} can be used to write down the following matrix relation,
\begin{align}
(U^{\rm T})_{i \alpha}^{\vphantom{*}} m_{\alpha \beta}^{\phantom{\rm T}}\,U_{\beta j}^{\phantom{\rm T}} =
-\left[M_I^{-1/2} \left(m_{\rm D}\right)_{I\alpha} U_{\alpha i}^{\phantom{\rm T}} \right]^{\rm T}\delta_{IJ}^{\phantom{\rm T}}\left[M_J^{-1/2} \left(m_{\rm D}\right)_{J\beta} U_{\beta j}^{\phantom{\rm T}}\right] = m_i\,\delta_{ij} \,, 
\end{align}
which can be solved for the Dirac mass matrix $\left(m_{\rm D}\right)_{I\alpha}$, or equivalently, for the Yukawa matrix $y_{I\alpha}$,
\begin{align}
\label{eq:CIP}
y_{I\alpha} = \frac{\left(m_{\rm D}\right)_{I\alpha}}{v/\sqrt{2}} = \frac{\ii}{v/\sqrt{2}}\,M_I^{1/2} R_{Ii}^{\phantom{\rm T}}\, m_i^{1/2} (U^\dagger)_{i\alpha}^{\vphantom{*}} \,,
\end{align}
This is nothing but the famous \textit{Casas-Ibarra parametrization} (CIP) of the RHN Yukawa matrix~\cite{Casas:2001sr}.
The matrix $R$ in Eq.~\eqref{eq:CIP} is a complex rotation matrix that satisfies $RR^{\rm T} = \mathbb{1}_{2\times2}$ (but $R^{\rm T}R \neq \mathbb{1}_{3\times3}$).
It can be parametrized in terms of a complex rotation angle $z\in\mathbb C$ and a discrete parameter $\zeta=\pm 1$, which distinguishes between a positive branch $R_+$ and a negative branch $R_-$ of possible $R$ matrices.
For NH and IH, we can respectively write the complex rotation matrix $R$ as follows,
\begin{align}
\textrm{NH:} \quad R_\zeta\left(z\right) & =
\begin{pmatrix}
0 & +\cos z & \zeta\sin z \\
0 & -\sin z & \zeta\cos z
\end{pmatrix} \,, &
\textrm{IH:} \quad R_\zeta\left(z\right) & =
\begin{pmatrix}
+\cos z & \zeta\sin z & 0\\
-\sin z & \zeta\cos z & 0
\end{pmatrix} \,.
\end{align}
The two real DOFs contained in $z = z_{\rm R}+\ii\,z_{\rm I}$ reflect the mismatch between high-energy and low-energy parameters in the 2RHN seesaw.%
\footnote{In the 3RHN seesaw, the matrix $R$ is correspondingly parametrized in terms of three complex angles, $z_{12}$, $z_{23}$, and $z_{13}$, which correspond to the six real DOFs that cannot be constrained by the low-energy data in this model.}
In our analysis in Sec.~\ref{sec:results}, we will use the CIP in Eq.~\eqref{eq:CIP} to scan over all possible flavor models that are consistent with the low-energy neutrino observables within their $3\,\sigma$ ranges (see Tab.~\ref{tab:data}).
In doing so, we will make use of several properties of the Yukawa couplings $y_{I\alpha}$ as functions of $z_{\rm R}$ and $z_{\rm I}$ in the limit of a negligibly small RHN mass splitting, $M_2 - M_1 \rightarrow 0$.
First of all, we note that shifting the real part of the complex rotation angle by $\Delta z_{\rm R} = \pi/2$ results in a row exchange as well as in a sign flip in the second row of the matrix $y_{I\alpha}$.
This leads to the following pattern when shifting $z_{\rm R}$ by $\Delta z_{\rm R} = \pi/2$ in four discrete steps,
\begin{align}
\label{eq:zRshift}
M_1 = M_2 \,,\quad
\begin{pmatrix}+y_{1\alpha}\\+y_{2\alpha}\end{pmatrix} \overset{\Delta z_{\rm R} = \pi/2}{\longrightarrow} \begin{pmatrix}+y_{2\alpha}\\-y_{1\alpha}\end{pmatrix} \overset{\Delta z_{\rm R} = \pi}{\longrightarrow} \begin{pmatrix}-y_{1\alpha}\\-y_{2\alpha}\end{pmatrix} \overset{\Delta z_{\rm R} = 3\pi/2}{\longrightarrow} \begin{pmatrix}-y_{2\alpha}\\+y_{1\alpha}\end{pmatrix} \overset{\Delta z_{\rm R} = 2\pi}{\longrightarrow}\begin{pmatrix}+y_{1\alpha}\\+y_{2\alpha}\end{pmatrix} \,.
\end{align}
All relevant quantities in our parameter study in Sec.~\ref{sec:results} will depend on products of at least two Yukawa couplings.
For this reason, it will be enough if we restrict ourselves to scanning over the parameter range $z_{\rm R} \in \left[0,\pi\right)$.
Next to Eq.~\eqref{eq:zRshift}, the CIP also leads to two other useful relations,
\begin{align}
\label{eq:reflections}
M_1 & = M_2 \,, & z_{\rm R} & \rightarrow \pi/2 - z_{\rm R} \,, & \left(\zeta,\delta,\sigma\right) & \rightarrow -\left(\zeta,\delta,\sigma\right) & \quad\Rightarrow\qquad \begin{pmatrix}+y_{1\alpha}\\+y_{2\alpha}\end{pmatrix} & \rightarrow \begin{pmatrix}+y_{2\alpha}^*\\+y_{1\alpha}^*\end{pmatrix} \,,  \\\nonumber
M_1 & = M_2 \,, & z_{\rm R} & \rightarrow \pi - z_{\rm R} \,, & \left(z_{\rm I},\zeta\right) & \rightarrow -\left(z_{\rm I},\zeta\right) & \quad\Rightarrow\qquad \begin{pmatrix}+y_{1\alpha}\\+y_{2\alpha}\end{pmatrix} & \rightarrow \begin{pmatrix}-y_{1\alpha}\\+y_{2\alpha}\end{pmatrix} \,.
\end{align}
These symmetry properties of the CIP will be directly visible in our numerical results in Sec.~\ref{sec:results}.
Finally, we remark that the Yukawa couplings $y_{I \alpha}$ grow exponentially as functions of $\left|z_{\rm I}\right|$ for $\left|z_{\rm I}\right| \gg 1$,
\begin{align}
\label{eq:zI}
\left|z_{\rm I}\right| \gg 1 \quad\Rightarrow\qquad y_{1\alpha} \approx \sign(z\w I)\,\ii\, y_{2\alpha} \propto e^{\left|z_{\rm I}\right|} \,.
\end{align}
By construction, the CIP still results in Yukawa couplings that are in accord with the small SM neutrino masses in this case.
However, for $\left|z_{\rm I}\right| \gg 1$, this can only be achieved at the cost of fine-tuned cancellations among the different entries in the RHN Yukawa matrix.
In the context of the neutrino option, there is no reason why such a situation should be realized.
In our parameter study, we will therefore restrict ourselves to the range $z_{\rm I} \in \left[-2,+2\right]$ to avoid fine-tuned RHN Yukawa couplings.


\subsection{Baryon asymmetry}
\label{subsec:bau}


The baryon asymmetry of the Universe is typically quantified in terms of the baryon-to-photon ratio in the present era, $\eta_B^0$.
The most precise value for $\eta_B^0$ follows from the observations of the temperature anisotropies in the cosmic microwave background by the PLANCK satellite~\cite{Aghanim:2018eyx},
\begin{align}
\label{eq:bau}
\eta_B^{\rm obs} = \frac{n_B^0}{n_\gamma^0} \simeq 6.1 \times 10^{-10} \left(\frac{h^2\Omega_B^0}{0.0224}\right) \,.
\end{align}
Here $n_B^0$ and $n_\gamma^0$ are the number densities of baryons and photons, respectively, and $h^2\Omega_B^0 \simeq 0.0224$ is a measure for the energy density contained in baryons.
In this section, we will now discuss how resonant leptogenesis in the type-I seesaw manages to reproduce the baryon asymmetry in Eq.~\eqref{eq:bau}.


A first-principles treatment of leptogenesis requires one to perform computations in nonequilibrium \textit{quantum field theory} (QFT)~\cite{Anisimov:2010dk}, which is beyond the scope of this work.
Instead, we will carry out a semianalytical analysis similar to the one in Ref.~\cite{Bambhaniya:2016rbb}, which also studies resonant leptogenesis in the 2RHN seesaw model.
We begin by relating $\eta_B^0$ to the three LH-lepton-doublet asymmetries $\eta_{L_{\alpha}}^{\rm lptg}$ that are produced by the various $L_{\alpha}$-number-violating interactions during leptogenesis,
\begin{align}
\label{eq:etaB0}
\eta_B^0 = \frac{g_{*,s}^0}{g_{*,s}}\,C_{\rm sph}\sum_{\alpha} \eta_{L_\alpha}^{\rm lptg} \,.
\end{align}
Here, $g_{*,s}^0$ and $g_{*,s}$ count the effective number of relativistic DOFs that contribute to the entropy density of the thermal bath, $s_{\rm rad} = 2\pi^2/45\,g_{*,s}\,T^3$, at the present time and at the time of leptogenesis, respectively.
In the following, we will use the usual SM values for these two quantities,
\begin{align}
g_{*,s}^0 = \frac{43}{11} \,, \quad g_{*,s} = \frac{427}{4} \qquad\Rightarrow\qquad \frac{g_{*,s}^0}{g_{*,s}} = \frac{172}{4697} \simeq \frac{1}{27.3} \,.
\end{align}
$C_{\rm sph}$ in Eq.~\eqref{eq:etaB0} accounts for the conversion from the LH-lepton-doublet asymmetries $\eta_{L_\alpha}^{\rm lptg}$ to $\eta_B$ by means of EW sphalerons.
Let us now compute $C_{\rm sph}$ based on the analysis in Ref.~\cite{Pilaftsis:2005rv}.
In doing so, we will neglect the effect of spectator processes~\cite{Barbieri:1999ma,Buchmuller:2001sr,Garbrecht:2014kda,Garbrecht:2019zaa} and make use of the fact that, at temperatures $T \sim 10^5\,\textrm{GeV}$, all SM interactions eventually reach chemical equilibrium~\cite{Bodeker:2019ajh}.


First, we note that the total lepton asymmetry $\eta_L$ receives contributions from both LH lepton doublets $L_\alpha$ and RH lepton singlets $\ell_\alpha^{\rm R}$. 
In chemical equilibrium, one finds the following relation,
\begin{align}
\label{eq:LLR}
\eta_{\ell_\alpha^{\rm R}} = \frac{1}{2}\,\eta_{L_\alpha} - \frac{2}{21}\,\eta_L^{\rm L} \qquad\Rightarrow\qquad \eta_L = \eta_L^{\rm L} + \eta_L^{\rm R} = \frac{17}{14}\,\eta_L^{\rm L} \,,\quad \eta_L^{\rm L} = \sum_\alpha \eta_{L_{\alpha}} \,,\quad \eta_L^{\rm R} = \sum_\alpha \eta_{\ell_\alpha^{\rm R}} \,.
\end{align}
Second, it is important to notice that the transport equations that describe the evolution of the LH-lepton-doublet asymmetries $\eta_{L_\alpha}$ typically do not contain an explicit collision operator accounting for the EW sphalerons processes, although EW sphalerons certainly do violate the $L_\alpha$ number densities.
In this case, one has to split the total lepton asymmetry $\eta_L$ into two contributions~\cite{Pilaftsis:2005rv},
\begin{align}
\eta_L = \eta_L^{\rm lptg} + \eta_L^{\rm sph} \,.
\end{align}
Here, $\eta_L^{\rm lptg}$ denotes the effective lepton asymmetry that is generated during leptogenesis and whose evolution follows from solving a coupled set of transport equations.
The asymmetry $\eta_L^{\rm sph}$, on the other hand, is induced by the interaction density of the EW sphalerons, $\gamma_{\rm sph}$.
The effective leptogenesis asymmetry $\eta_L^{\rm lptg}$ can be related to the total baryon and lepton asymmetries, $\eta_L$ and $\eta_B$,~\cite{Pilaftsis:2005rv}
\begin{align}
\eta_L^{\rm lptg} = \eta_L - \eta_B = \eta_{L-B} = - \eta_{B-L} \,.
\end{align}
This reflects the well-known fact that the lepton number generated during leptogenesis should actually be regarded as a negative $B\!-\!L$ charge whose $B$ component happens to be zero.
The total $B$, $L$, and $B-L$ asymmetries are related to each other by standard factors, such that
\begin{align}
\label{eq:BLBL}
\eta_B = -\frac{28}{51}\,\eta_L = \frac{28}{79}\,\eta_{B-L} = -\frac{28}{79}\,\eta_L^{\rm lptg} \,.
\end{align}
Eq.~\eqref{eq:LLR} also applies at the level of the individual lepton asymmetries $\eta_L^{\rm lptg}$ and $\eta_L^{\rm sph}$~\cite{Pilaftsis:2005rv}.
Combining our results in Eqs.~\eqref{eq:LLR} and \eqref{eq:BLBL}, we therefore obtain the following sphaleron conversion factor,
\begin{align}
\label{eq:Csph}
\eta_B = C_{\rm sph}\sum_{\alpha} \eta_{L_\alpha}^{\rm lptg} \,,\quad C_{\rm sph} = -\frac{17}{14}\times\frac{28}{79} = -\frac{34}{79} \simeq -0.43 \,.
\end{align}
It is interesting to compare this results with other expressions for $C_{\rm sph}$ that one frequently encounters in the literature.
In our notation, we are also able to write down the following three relations,
\begin{align}
\eta_B = -\frac{28}{79}\,\eta_{L-B} \,,\quad \eta_B = -\frac{28}{51}\,\eta_L \,,\quad \eta_B = -\frac{2}{3}\,\eta_L^{\rm L} \,.
\end{align}
None of these relations should be confused with the correct relation in Eq.~\eqref{eq:Csph}.%
\footnote{The sphaleron factor in Eq.~\eqref{eq:Csph} can also be estimated by consistently neglecting the effect of the RH lepton singlets.
In this case, one can simply write $\eta_B = -2/3\,\eta_L^{\rm L} \approx -2/3\,\eta_L$, such that $\eta_B \approx -2/5\,\eta_{L-B} = -2/5\,\eta_L^{\rm lptg} \approx -2/5\,\sum_\alpha \eta_{L_\alpha}^{\rm lptg}$.
This estimate results in $C_{\rm sph} \approx -2/5 = -0.40$, which deviates from our result by less than $10\,\%$.
In the following, we will, however, stick to our result $C_{\rm sph} = -34/79$.
We thank Daniele Teresi for a helpful discussion on this point.}


We anticipate the temperature scale of leptogenesis to be $T \sim 10^6\cdots10^7\,\textrm{GeV}$ in our scenario.
In this case, the $\mu$ and $\tau$ Yukawa interactions have already reached chemical equilibrium during leptogenesis, such that we do not need to worry about coherence\,/\,decoherence effects in flavor space.
Therefore, instead of solving matrix equations for quantum-mechanical density matrices, it is sufficient to restrict oneself to semiclassical Boltzmann equations for number densities.
In this paper, we will follow the analysis in Ref.~\cite{Deppisch:2010fr} (see also Ref.~\cite{Bambhaniya:2016rbb}), which studies the evolution of the LH-lepton-doublet asymmetries $\eta_{L_\alpha}^{\rm lptg}$ during resonant leptogenesis based on a set of flavored Boltzmann equations.
The final asymmetries at the end of leptogenesis can then be written as follows~\cite{Deppisch:2010fr,Bambhaniya:2016rbb},
\begin{align}
\label{eq:etaLlepto}
\eta_{L_\alpha}^{\rm lptg} = \frac{3}{2}\sum_{I} \frac{\varepsilon_{I\alpha}}{K_\alpha^{\rm eff} z_\alpha^{\phantom{\rm eff}}} \,.
\end{align}
In writing down this expression, we assumed that leptogenesis always terminates before the EW sphalerons fall out of thermal equilibrium at temperatures around $T_{\rm sph} \simeq 130\,\textrm{GeV}$~\cite{DOnofrio:2014rug}.
Given that we are interested in leptogenesis at temperatures far above the EW scale, $T \gg T_{\rm sph}$, this assumption is always satisfied.
Another assumption entering Eq.~\eqref{eq:etaLlepto} is that one of the $\eta_{L_\alpha}^{\rm lptg}$ asymmetries should be larger than the two others, such that leptogenesis is predominantly driven by the asymmetry in one specific lepton flavor.
We performed an explicit numerical analysis of the Boltzmann equations in Ref.~\cite{Deppisch:2010fr} to confirm that this assumption is justified in our scenario to first approximation.
We also confirmed that the semianalytical expression in Eq.~\eqref{eq:etaLlepto} allows one to reproduce the fully numerical result with a precision of $\Order(10\%)$.
For these reasons, we will restrict ourselves to working with Eq.~\eqref{eq:etaLlepto} in the following.
For more details on the Boltzmann equations and the derivation of Eq.~\eqref{eq:etaLlepto}, we refer the reader to the detailed presentation in Ref.~\cite{Deppisch:2010fr}.


Let us now spell out the meaning of $\varepsilon_{I\alpha}$, $K_\alpha^{\rm eff}$, and $z_\alpha$ in Eq.~\eqref{eq:etaLlepto}. 
The parameter $z_\alpha$ quantifies the point in time when the lepton asymmetry $\eta_{L_\alpha}^{\rm lptg}$ ceases to evolve at the end of leptogenesis,
\begin{align}
z_\alpha \simeq 1.25\,\ln\left(25\,K_\alpha^{\rm eff}\right) \,.
\end{align}
$K_\alpha^{\rm eff}$ is an effective measure for the strength of processes that wash out the asymmetry in the $\alpha$-flavor channel.
Large values, $K_\alpha^{\rm eff} \gg 1$, correspond to strong washout, while $K_\alpha^{\rm eff} \sim 1$ corresponds to weak washout.
The parameter $K_\alpha^{\rm eff}$ is defined in terms of the standard RHN decay parameters $K_I$,
\begin{align}
\label{eq:Kaeff}
K_\alpha^{\rm eff} = \kappa_\alpha \sum_I K_I \mathcal{B}_{I\alpha} \,,\quad K_I = \frac{\Gamma_I}{\zeta\left(3\right)H\left(T = M_I\right)} \,, \quad \mathcal{B}_{I\alpha} = \frac{\left|y_{I\alpha}\right|^2}{\left(yy^\dagger\right)_{II}} \,.
\end{align}
Here, $\mathcal{B}_{I\alpha}$ denotes the branching ratio of $N_I$ decays into LH lepton doublets of flavor $\alpha$ at tree level, and $H\left(T = M_I\right)$ is the Hubble rate evaluated at a temperature equal to the RHN mass $M_I$,
\begin{align}
H\left(T = M_I\right) = \left(\frac{\pi^2\,g_{*,\rho}}{90}\right)^{1/2} \frac{M_I^2}{M_{\rm Pl}} \,,
\end{align}
where $g_{*,\rho} = 427/4$ is the effective number of relativistic DOFs that contribute to the energy density of the thermal SM plasma, $\rho_{\rm rad} = \pi^2/30\,g_{*,\rho}\,T^4$, at the time of leptogenesis, and where we employed the reduced Planck mass $M_{\rm Pl} = \left(8\pi G\right)^{-1/2}\simeq 2.44\times10^{18}\,\textrm{GeV}$ (with $G$ being Newton's constant).
$\Gamma_I$ in Eq.~\eqref{eq:Kaeff} is the total $N_I$ tree-level decay rate at zero temperature,
\begin{align}
\Gamma_I = \Gamma\left(N_I\rightarrow L_\alpha + \phi\right) + \Gamma\left(N_I\rightarrow L_\alpha^{\rm C} + \phi^*\right) = \frac{(yy^\dagger)_{II}}{8\pi}\,M_I \,.
\end{align}
These definitions illustrate that the RHN decay parameter $K_I$ characterizes how strongly the RH neutrino $N_I$ is coupled to the thermal bath. 
For $K_I \gg 1$, the RHN interactions have (nearly) reached thermal equilibrium, for $K_I \ll 1$, 
they are far away from thermal equilibrium.
Eq.~\eqref{eq:Kaeff} also shows that $\sum_I K_I \mathcal{B}_{I\alpha} = K_\alpha$ can be regarded as the equivalent of $K_I$ in lepton flavor space.
The washout parameter $K_\alpha^{\rm eff} = \kappa_\alpha K_\alpha$, finally, is a rescaled version of $K_\alpha$, where the factor $\kappa_\alpha$ accounts for the effect of lepton-number-violating and lepton-flavor-violating two-to-two scattering processes,%
\footnote{In Ref.~\cite{Deppisch:2010fr}, the rescaling factor $\kappa_\alpha$ is expressed in terms of resummed Yukawa couplings $\bar{y}_{I\alpha}$ instead of the ordinary tree-level Yukawa couplings $y_{I\alpha}$.
The relation between these two sets of Yukawa couplings in the case of only two RH neutrinos can be found in Refs.~\cite{Dev:2014laa,Dev:2015wpa}.
We checked that replacing $y_{I\alpha}$ by $\bar{y}_{I\alpha}$ in our parameter study only leads to numerically insignificant changes.
For this reason, we decide to ignore this subtlety in the following.
We also point out that Eq.~\eqref{eq:kappa} agrees with the result in Ref.~\cite{Deppisch:2010fr} after substituting $\bar{y}_{I\alpha}\rightarrow y_{I\alpha}$, whereas it does not agree with the expression in Ref.~\cite{Bambhaniya:2016rbb}, which involves a number of typos.
We thank Bhupal Dev for a helpful discussion on this point.}
\begin{align}
\label{eq:kappa}
\kappa_\alpha  & = 2\sum_{I,J} \frac{\textrm{Re}\big[y_{I\alpha}^{\phantom{*}}y_{J\alpha}^*\big]\textrm{Re}\left[\left(yy^\dagger\right)_{IJ}\right] - \textrm{Im}\big[y_{I\alpha}^{\phantom{*}}y_{J\alpha}^*\big]\textrm{Im}\big[y_{I\alpha}^{\phantom{*}}y_{J\alpha}^*\big]}{\left(y^\dagger y\right)_{\alpha\alpha}\left[\left(yy^\dagger\right)_{II} + \left(yy^\dagger\right)_{JJ}\right]}\left(1-2\ii\,\frac{M_I-M_J}{\Gamma_I+\Gamma_J}\right)^{-1} \\\nonumber
& = 1 + 4\,\frac{\textrm{Re}\big[y_{1\alpha}^{\phantom{*}}y_{2\alpha}^*\big]\textrm{Re}\left[\left(yy^\dagger\right)_{12}\right] - \textrm{Im}\big[y_{1\alpha}^{\phantom{*}}y_{2\alpha}^*\big]\textrm{Im}\big[y_{1\alpha}^{\phantom{*}}y_{2\alpha}^*\big]}{\left(y^\dagger y\right)_{\alpha\alpha}\left[\left(yy^\dagger\right)_{11} + \left(yy^\dagger\right)_{22}\right]} \frac{\left(\Gamma_1 + \Gamma_2\right)^2}{\left(\Gamma_1 + \Gamma_2\right)^2 + 4\left(M_2 - M_1\right)^2}\,.
\end{align}
In the scattering operators leading to this expression for $\kappa_\alpha$, the contributions from on-shell RH neutrinos in the intermediate state have been subtracted.
These contributions are already accounted for by the decay and inverse-decay operators in the Boltzmann equations and must not be counted twice.
In our parameter scan in Sec.~\ref{sec:results}, we will mostly be interested in small RHN decay widths, $\Gamma_{1,2} \ll M_2-M_1$.
In this case, the rescaling factor $\kappa_\alpha$ typically obtains values close to one, $\kappa_\alpha \approx 1$.


The final ingredient in Eq.~\eqref{eq:etaLlepto} is the $CP$ asymmetry parameter $\varepsilon_{I\alpha}$, which represents the amount of $CP$ asymmetry that can be generated per RHN decay.
It is defined in terms of the partial decay widths $\Gamma_{1\ell}\left(N_I\rightarrow L_\alpha + \phi\right)$ and $\Gamma_{1\ell}\left(N_I\rightarrow L_\alpha^{\rm C} + \phi^*\right)$, which involve the RHN decay amplitudes at tree level as well as the radiative RHN \textit{vertex} (v) and \textit{self-energy} (s) corrections at one loop~\cite{Covi:1996wh},
\begin{align}
\varepsilon_{I\alpha} = \varepsilon_{I\alpha}^{\rm (v)} + \varepsilon_{I\alpha}^{\rm (s)} = \frac{\Gamma_{1\ell}\left(N_I\rightarrow L_\alpha + \phi\right) - \Gamma_{1\ell}\left(N_I\rightarrow L_\alpha^{\rm C} + \phi^*\right)}{\Gamma_I} \,.
\end{align}
The vertex contribution $\varepsilon_{I\alpha}^{\rm (v)} $ (also referred to as the $\varepsilon'$ or direct $CP$ asymmetry) reads,
\begin{align}
\label{eq:epsilonV}
\varepsilon_{I\alpha}^{\rm (v)} = \sum_{J \neq I} \frac{\Im\left[y^\ph_{I\alpha}y^*_{J\alpha}\yydag_{IJ}\right]} {\yydag_{II}\yydag_{JJ}} \frac{\Gamma_J}{M_I}\left[1 - \left(1 + \frac{M_J^2}{M_I^2}\right) \ln\left(1 + \frac{M_I^2}{M_J^2}\right)\right] \,,
\end{align}
while one finds for the self-energy contribution (also referred to as the $\varepsilon$ or indirect $CP$ asymmetry),
\begin{align}
\label{eq:epsilonS}
\varepsilon_{I\alpha}^{\rm (s)} = \sum_{J \neq I} \left\{\frac{\Im\left[y^\ph_{I\alpha} y^*_{J\alpha} \yydag_{IJ}\right]}{\yydag_{II}\yydag_{JJ}} + \frac{M_I}{M_J}\frac{\Im\left[y^\ph_{I\alpha} y^*_{J\alpha}\yydag_{JI}\right]}{\yydag_{II}\yydag_{JJ}}\right\} f_{IJ} \,,
\end{align}
where the function $f_{IJ} \propto 1/\left(M_I^2 - M_J^2\right)$ originates from the $N_J$ propagator in the $N_I$ decay diagram.


Resonant leptogenesis is based on the observation that $\varepsilon_{I\alpha}^{\rm (s)}$ can be resonantly enhanced in the case of a small RHN mass splitting, $\Delta M = M_2 - M_1 \ll M_1$, such that $f_{IJ}$ becomes exceptionally large.
In fact, given the naive semiclassical estimate $f_{IJ} \propto 1/\left(M_I^2 - M_J^2\right)$, one finds that $\varepsilon_{I\alpha}^{\rm (s)}$ appears to diverge for a vanishing mass splitting, $\Delta M \rightarrow 0$.
This, however, is an unphysical effect that reflects the breakdown of the semiclassical approximation.
A pair of exactly degenerate RH Majorana neutrinos represents, in fact, a single Dirac neutrino, such that lepton number remains conserved and the $CP$ asymmetry parameter identically vanishes, $\varepsilon_{I\alpha} = 0$.
This simple argument indicates that the function $f_{IJ}$ needs to be regularized in quantum fields theory,
\begin{align}
\frac{1}{M_I^2 - M_J^2} \:\:\rightarrow\:\: \frac{M_I^2 - M_J^2}{\left(M_I^2 - M_J^2\right)^2 + R_{IJ}} \,,
\end{align}
in order to avoid the singular behavior in the limit $\Delta M \rightarrow 0$.
The proper form of the regulator $R_{IJ}$ has been the subject of a long debate in the literature that has not yet been fully settled (see, \textit{e.g.}, \cite{Dev:2014laa,Dev:2015wpa,Iso:2013lba,Iso:2014afa,Garbrecht:2014aga,Dev:2014wsa,Kartavtsev:2015vto}).
In this paper, we do not have anything new to add to this debate.
Instead, we will simply adopt the results in Refs.~\cite{Dev:2014laa,Dev:2015wpa,Dev:2014wsa}, which managed to reproduce the same form of the regulator both in an analysis based on quantum-mechanical density matrix equations~\cite{Dev:2014laa,Dev:2015wpa} and a full QFT analysis based on Kadanoff--Baym equations~\cite{Dev:2014wsa}.
The main conclusion of Refs.~\cite{Dev:2014laa,Dev:2015wpa,Dev:2014wsa} is that the function $f_{IJ}$ actually receives two contributions of similar magnitude,
\begin{align}
\label{eq:foscmix}
f_{IJ} = f_{IJ}^{\rm osc} + f_{IJ}^{\rm mix} \,, \quad f_{IJ}^{\rm osc} = \frac{\left(M_I^2 - M_J^2\right) M_I \Gamma_J}{\left(M_I^2 - M_J^2\right)^2 + R_{IJ}^{\rm osc}} \,, \quad f_{IJ}^{\rm mix} = \frac{\left(M_I^2 - M_J^2\right) M_I \Gamma_J} {\left(M_I^2 - M_J^2\right)^2 + R_{IJ}^{\rm mix}} \,,
\end{align}
where $f_{IJ}^{\rm osc}$ and $f_{IJ}^{\rm mix}$ account for the contributions to the $CP$ asymmetry parameter $\varepsilon_{I\alpha}^{\rm (s)}$ from RHN flavor oscillations and RHN mixing, respectively.
If one decided to omit one these two contributions, the $CP$ asymmetry parameter $\varepsilon_{I\alpha}^{\rm (s)}$ would roughly decrease by a factor $2$.
This should be regarded as an upper bound on the theoretical uncertainty of our expression for $\varepsilon_{I\alpha}^{\rm (s)}$ in Eq.~\eqref{eq:epsilonS}.
The regulators $R_{IJ}^{\rm osc}$ and $R_{IJ}^{\rm mix}$ in Eq.~\eqref{eq:foscmix} are given by the following expressions,
\begin{align}
\label{eq:RIJoscmix}
R_{IJ}^{\rm osc} = \left(M_I \Gamma_I + M_J \Gamma_J\right)^2 \frac{\det\left[\Re\yydag\right]}{\yydag_{II}\yydag_{JJ}},\qquad R_{IJ}^{\rm mix} = M_I^2\,\Gamma_J^2 \,.
\end{align}
This concludes our discussion of the different ingredients that are necessary to compute the final baryon asymmetry in our scenario.
Combining all of the above results, we are now able to write
\begin{align}
\label{eq:etaB0final}
\eta_B^0 \simeq C \sum_{I,\alpha}\frac{\varepsilon_{I\alpha}}{1.25\,\ln\left(25\,K_\alpha^{\rm eff}\right)K_\alpha^{\rm eff}} \,,\quad C = -\frac{8772}{371063} \simeq -2.36 \times 10^{-2} \,,
\end{align}
which is the expression for $\eta_B^0$ that we will use to calculate the baryon asymmetry in Sec.~\ref{sec:results}.


\subsection{Electroweak scale}
\label{subsec:higgs}


Let us now turn to the generation of the EW scale in the type-I seesaw.
In the pure Standard Model without RH neutrinos, the Higgs doublet $\phi$ possesses the following scalar potential,
\begin{align}
\label{eq:VHiggs}
V_{\rm SM} = -\mu^2 \left|\phi\right|^2 + \lambda\left|\phi\right|^4 \,, \quad \left|\phi\right|^2 = \phi^\dagger\phi = \phi_+\phi_- + \phi_0^*\,\phi_0^{\phantom{*}} \,.
\end{align}
Here, $\mu$ denotes the Higgs mass parameter, and $\lambda$ is the quartic Higgs self-coupling.
The sign of the mass term in Eq.~\eqref{eq:VHiggs} is chosen such that a real and positive mass parameter, $\mu > 0$, results in EWSB.
The Standard Model is based on the assumption that $\mu \sim 100\,\textrm{GeV} > 0$ around a renormalization scale $Q \sim 100\,\textrm{GeV}$; but it does not offer any intrinsic justification for this assumption.
In the true vacuum after EWSB, the Higgs field has a nonzero VEV $\sqrt{2}\left<\phi_0\right> = v = \mu/\sqrt{\lambda}$, and the physical Higgs boson $h_0$ possesses a mass $m_{h_0} = \sqrt{2}\,\mu = \sqrt{2\lambda}\,v$.
We emphasize that $\mu$ is the only explicit mass scale in the Standard Model.
Once we set $\mu\rightarrow0$, the SM Lagrangian becomes scale-invariant.%
\footnote{The scale of \textit{quantum chromodynamics} (QCD), $\Lambda_{\rm QCD} \sim 100\,\textrm{MeV}$, is not an explicit input scale, but generated via dimensional transmutation by the \textit{renormalization group} (RG) running of the strong gauge coupling constant.}


Motivated by this observation, the Majorana-neutrino option is based on the assumption that the Higgs potential satisfies scale-invariant boundary conditions above the RHN mass thresholds,
\begin{align}
\label{eq:VUV}
Q > \textrm{max}\left\{M_I\right\} \quad\Rightarrow\quad V_{\rm UV} = \lambda\left|\phi\right|^4 \,.
\end{align}
The Higgs mass parameter $\mu$ necessary for EWSB is then induced by RHN threshold corrections to the Higgs potential, $\Delta V \supset -\mu^2 \left|\phi\right|^2$, when matching the full theory (including dynamical RH neutrinos) at high energies onto the \textit{effective field theory} (EFT) (without dynamical RH neutrinos) at low energies.
In the case of two nearly degenerate RH neutrinos, this matching is done at $Q_0 = M_1 \simeq M_2$, which is the energy scale at which both RH neutrinos decouple~\cite{Appelquist:1974tg}.
The threshold corrections $\Delta V$ are encoded in the one-loop effective \textit{Coleman--Weinberg} (CW) potential~\cite{Coleman:1973jx},
\begin{align}
\label{eq:VCW}
\Delta V = V_{\rm UV} - V_{\rm EFT} = \frac{\left(-2\right)}{64\pi^2}\sum_I M_I^4\left(\phi\right)\ln\frac{M_I^2\left(\phi\right)}{Q^2} \,.
\end{align}
The CW formula can be derived by computing one-loop vacuum (bubble) diagrams, using dimensional regularization to keep track of the infinities arising in the one-loop momentum integrals.
These infinities are canceled by an appropriate set of counterterms in the \MSbar renormalization scheme.
In our notation, the \MSbar renormalization scale $\bar{Q}$ is given by $\bar{Q} = e^{-3/4}\,Q$.
In this sense, the scale $Q$ can be regarded as the renormalization scale in a different scheme that deviates from the \MSbar scheme by a finite shift along the RG flow.
The advantage of employing $Q$ rather than $\bar{Q}$ in Eq.~\eqref{eq:VCW} is that, in this scheme, the CW potential no longer contains nonlogarithmic terms that do not depend on the renormalization scale; in Eq.~\eqref{eq:VCW}, all terms are proportional to logarithms of the form $\ln M_I^2/Q^2$.
This allows one to minimize the absolute value of the CW potential by setting $Q$ to the typical mass scale of the theory.
In our analysis, we will consequently fix the renormalization scale at the RHN decoupling scale, $Q_0 = M_1$, corresponding to $\bar{Q}_0 = e^{-3/4}\,M_1$ in the \MSbar scheme.
In this way, we intend to remove large logarithms from the CW potential, which
we expect to improve the quality of the perturbative series.
This is in line with the discussion in Refs.~\cite{Casas:1998cf,Casas:1999cd}, which states that the choice $Q_0 = M_1$ amounts to a leading-log resummation to all orders in perturbation theory that minimizes the $Q$ dependence of the CW potential.
In future work, it would be interesting to confirm these statements by an explicit calculation of higher-order corrections in the type-I seesaw model.
In the remainder of this paper, we will, however, simply stick to our choice $Q_0 = M_1$, cautioning that the scheme dependence of the EFT matching analysis requires further scrutiny. 


The $\left(-2\right)$ prefactor in Eq.~\eqref{eq:VCW} follows from the Fermi-Dirac statistics of the RH neutrinos that run in the one-loop vacuum diagrams.
In our renormalization scheme, the negative sign of this factor eventually leads to a negative sign of the Higgs mass term at the matching scale $Q_0 = M_1$.
This illustrates that the type-I seesaw is capable of explaining the tachyonic mass term in the Higgs potential that is responsible for EWSB at low energies.
The $\phi$-dependent masses $M_I\left(\phi\right)$ in Eq.~\eqref{eq:VCW} represent the two large mass eigenvalues that one finds when performing a Autonne-Takagi factorization of the total neutrino mass matrix $\mathcal{M}$ in Eq.~\eqref{eq:M} after replacing the Higgs VEV $v$ by a classical and homogeneous Higgs field background $\phi$.
Again, similarly as in Sec.~\ref{subsec:neutrinos}, the total neutrino mass matrix can be approximately diagonalized by performing a perturbative expansion in powers of the Higgs field.
While we were only interested in the leading-order result in Sec.~\ref{subsec:neutrinos} [see Eq.~\eqref{eq:Mblock}], we now have to compute the squared masses $M_I^2\left(\phi\right)$ up to $\Order(\phi^4)$, in order to identify the threshold corrections to the quadratic and quartic terms in the Higgs potential.
This procedure allows us to write the threshold corrections $\Delta V$ as a power series in the Higgs doublet $\phi$,
\begin{align}
\Delta V = \Delta V_0 - \Delta \mu^2 \left|\phi\right|^2 + \Delta\lambda \left|\phi\right|^4 + \Order\left(|\phi|^6\right) \,,
\end{align}
where $\Delta V_0$ denotes a constant contribution to the vacuum energy that we can ignore for our purposes.


We obtain the following expressions for the threshold corrections $\Delta\mu^2$ and $\Delta\lambda$,
\begin{align}
\label{eq:Deltaml}
\Delta\mu^2 & = \frac{1}{8\pi^2} \sum_I\left[\left(yy^\dagger\right)_{II}\,M_I^2\left(\ln\frac{M_I^2}{Q^2} + \frac{1}{2}\right)\right] \,, \\\nonumber
\Delta\lambda & = \frac{1}{8\pi^2} \sum_{I\neq J}\left[\left(yy^\dagger\right)_{II}^2 L_{IJ}^{(1)} - \frac{1}{2}\pTr{yy^\dagger y^*y^{\rm T}} L_{IJ}^{(2)} + \frac{1}{2}\pTr{yy^\dagger yy^\dagger} L_{IJ}^{(3)}\right] \,,
\end{align}
where the three loop functions $L_{IJ}^{(1)}$, $L_{IJ}^{(2)}$, and $L_{IJ}^{(3)}$ are defined as follows,
\begin{align}
L_{IJ}^{(1)} &= \frac{M_J}{M_J-M_I}\,\ln\frac{M_J}{M_I} - 1 \,,
&
L_{IJ}^{(2)} &= \frac{M_I M_J}{M_J^2-M_I^2}\,\ln\frac{M_J}{M_I} \,,
&
L_{IJ}^{(3)} &= \frac{M_I^2}{M_J^2-M_I^2}\,\ln\frac{M_I^2}{Q^2} - \frac{1}{4} \,.
\end{align}
In the limit of a small RHN mass splitting, $M_2 \simeq M_1$, these expressions can be simplified to
\begin{align}
\Delta\mu^2 &\simeq \frac{\pTr{yy^\dagger}}{8\pi^2}\,M_1^2\left(\ln\frac{M_1^2}{Q^2} + \frac{1}{2}\right) \,, \quad
&
\Delta\lambda &\simeq -\frac{\pTr{yy^\dagger y^*y^{\rm T}}}{16\pi^2} -\frac{\pTr{yy^\dagger yy^\dagger}}{16\pi^2}\left(\ln\frac{M_1^2}{Q^2} + \frac{3}{2}\right)  \,.
\end{align}
In particular, we obtain the following threshold corrections at the RHN decoupling scale $Q_0 = M_1$,
\begin{align}
\Delta\mu_0^2 &= \left.\Delta\mu^2\right|_{Q_0} \simeq \frac{\pTr{yy^\dagger}}{16\pi^2}\,M_1^2 \,,
&
\Delta\lambda_0 &= \left.\Delta\lambda^{\vphantom{2}}\right|_{Q_0} \simeq -\frac{1}{16\pi^2}\left[\pTr{yy^\dagger y^*y^{\rm T}} + \frac{3}{2}\pTr{yy^\dagger yy^\dagger}\right] \,.
\end{align}
These results conform with the naive expectations $\Delta\mu_0^2  \sim y^2/\left(16\pi^2\right)M_1^2$ and $\Delta\lambda_0 \sim y^4/\left(16\pi^2\right)$.
We observe that the threshold correction to the Higgs mass parameter turns out to have a positive sign, $\Delta\mu_0^2 > 0$, which is the sign needed to induce EWSB at low energies.
The threshold correction to the quartic Higgs self-coupling, on the other hand, turns out to have a negative sign, $\Delta\lambda_0 < 0$, which may raise the concern of a vacuum instability in the Higgs potential.
However, at this point, it is important to remember that the Higgs potential already contains a quartic coupling at tree level [see Eq.~\eqref{eq:VUV}].
Compared to this tree-level coupling, $\Delta\lambda_0$ is typically tiny, given that it is suppressed by the combination of four Yukawa couplings and a loop factor.
The effect of the threshold correction $\Delta\lambda_0$ can therefore be easily compensated for by a shift in the tree-level coupling.
This allows us to neglect $\Delta\lambda_0$ and focus on the matching of the Higgs mass parameter in the following.


\begin{table}[t]
\begin{center}
\begin{tabular}[t]{c|S[table-format=1.5]|S[table-format=1.5]}
\toprule
          & {$\bar Q=173.2\GeV$} & {$\bar Q=10^7\GeV$} \\
\midrule
$g_1$     & 0.4639             & 0.4957            \\
$g_2$     & 0.6476             & 0.5961            \\
$g_3$     & 1.167              & 0.7614            \\
$y_t$     & 0.9503             & 0.6400            \\
$y_b$     & 0.02401            & 0.01434           \\
$y_\tau$  & 0.01020            & 0.01053           \\
$\lambda$ & 0.1276             & 0.01173           \\
$\mu$     & {\SI{93.54}{GeV}}          & {\SI{101.7}{GeV}}         \\
\bottomrule
\end{tabular}
\caption{List of SM parameters relevant for our RGE analysis: $g_i$ are the SM gauge couplings (where $g_1=\sqrt{5/3}\,g_Y$), while $y_t$, $y_b$, and $y_\tau$ denote the Yukawa couplings of the top quark, bottom quark, and tau lepton, respectively.
The second column contains the input values for these parameters at the top-quark mass scale, $\bar Q=173.2\GeV$ (see text for details), while the third column shows the computed values at $\bar Q=10^7\GeV$ based on our numerical solution of the two-loop RG equations.}
\label{tab:rges}
\end{center}
\end{table}


\begin{figure}
\begin{center}
\includegraphics[width=0.67\textwidth]{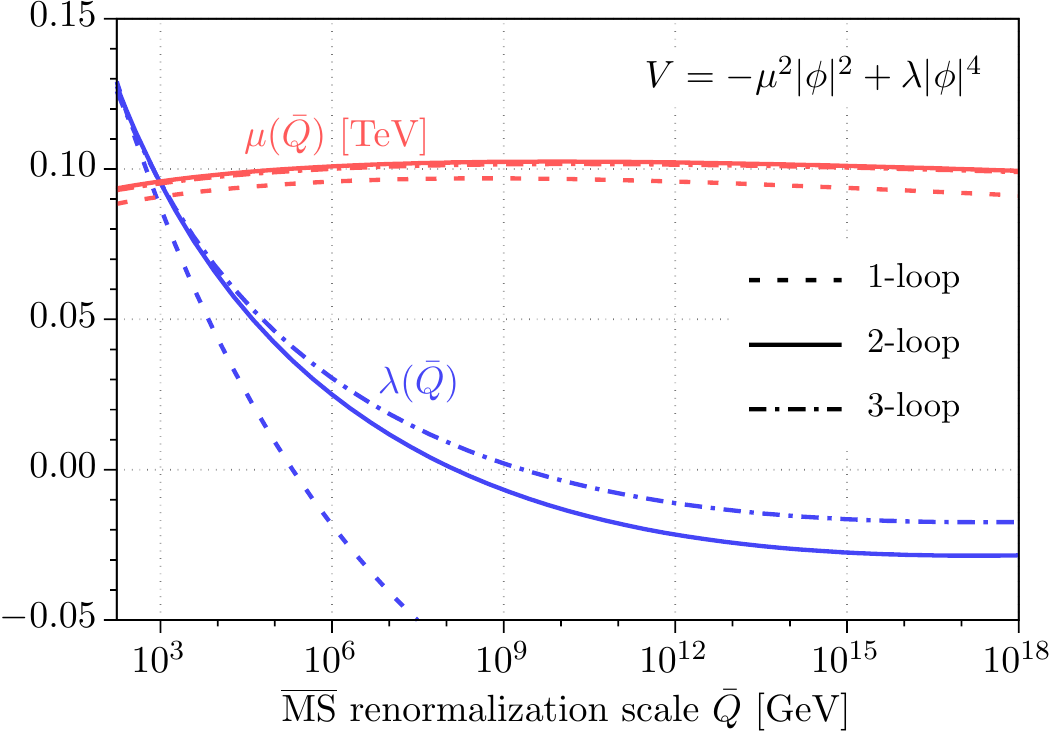}
\caption{RG running of the Higgs mass parameter $\mu$ and quartic Higgs self-coupling constant $\lambda$ as functions of the \MSbar renormalization scale $\bar{Q} = e^{-3/4}\,Q$ in the Standard Model.
For the purposes of this paper, we are only interested in $\bar{Q}$ values of at most $\mathcal{O}\left(10^7\right)\,\textrm{GeV}$.
The metastability\,/\,instability of the EW vacuum at higher energies needs to be addressed by new physics beyond the type-I seesaw.}
\label{fig:rge}
\end{center}
\end{figure}


At the RHN decoupling scale $Q_0 = M_1$, the type-I seesaw can be matched onto the \textit{SM effective field theory} (SMEFT)~(see Ref.~\cite{Brivio:2017vri} for a review).
A complete one-loop matching of the two theories including operators up to dimension five on the SMEFT side has been performed in Ref.~\cite{Brivio:2018rzm}.
There, it has been shown that the dominant outcome of the one-loop matching consists of (i) the threshold corrections to the Higgs potential in Eq.~\eqref{eq:Deltaml} and (ii) nonzero Wilson coefficients for the dimension-five Weinberg operator~\cite{Weinberg:1979sa}, which results in the SM neutrino masses after EWSB.
The one-loop Wilson coefficients computed in Ref.~\cite{Brivio:2018rzm} also capture radiative corrections to the tree-level neutrino mass matrix $m_{\alpha\beta}$ in Eq.~\eqref{eq:mab}.
However, these corrections only become relevant in the case of large (and fine-tuned) Yukawa couplings, \textit{i.e.}, for large absolute values of the parameter $z_{\rm I}$ [see Eq.~\eqref{eq:zI}].
In our parameter study in Sec.~\ref{sec:results}, we will not be interested in this regime.
This leaves the threshold correction to the Higgs mass parameter, $\Delta\mu_0^2$, as the only quantity that we need to explicitly account for, for the purposes of the matching analysis in this paper.
On the SMEFT side, the RG running of the Higgs mass parameter $\mu^2$ is controlled by the standard \textit{RG equations} (RGEs) of the Standard Model.%
\footnote{The RG running of the Wilson coefficients of the dimension-five Weinberg operator is numerically insignificant~\cite{Brivio:2018rzm}.
This is the SMEFT version of the statement that the RG running of the type-I seesaw parameters can typically be neglected~\cite{Antusch:2005gp,Brdar:2015jwo,Bambhaniya:2016rbb}.
For this reason, we will not consider any RGEs in the BSM sector of our model.}
Our matching condition therefore amounts to the requirement that the running Higgs mass parameter $\mu^2\left(Q\right)$ must equal the threshold correction $\Delta\mu_0^2$ at the RHN decoupling scale,
\begin{align}
\label{eq:matching}
\mu_0^2 = \mu^2\left(Q_0 = M_1\right) = \frac{\pTr{yy^\dagger}}{16\pi^2}\,M_1^2 = \Delta\mu_0^2\,.
\end{align}


In order to determine the value of $\mu^2\left(Q_0\right)$ on the left-hand side of Eq.~\eqref{eq:matching}, we need to solve the RG equations of the Standard Model, which we employ from Ref.~\cite{Buttazzo:2013uya}.
The outcome of our analysis is shown in Fig.~\ref{fig:rge}, alongside our results for the running quartic Higgs self-coupling.
In a first step, we solve the RG equations in the \MSbar renormalization scheme at one-loop, two-loop, and three-loop order~\cite{Buttazzo:2013uya}.
As evident from Fig.~\ref{fig:rge}, the result of the three-loop analysis for $\mu\left(\bar{Q}\right)$ does not significantly improve over the result of the two-loop analysis.
In the following, we will therefore restrict ourselves to working with the results of the two-loop RGEs, in combination with the one-loop threshold corrections to the running SM parameters at the top-quark mass scale, $\bar{Q} = m_t$, in order to fix the initial conditions of the two-loop RGEs.
These initial conditions are given in Tab.~\ref{tab:rges}, where we also show the corresponding values at the typical RHN scale of $10^7$ GeV, which we obtain after numerically solving the two-loop RGEs.
At values of the renormalization scale relevant for the Majorana-neutrino option, $\bar{Q} \sim 10^6\cdots10^7\,\textrm{GeV}$, the Higgs mass parameter turns out to exhibit only a very weak dependence on the renormalization scale (see Fig.~\ref{fig:rge}).
Over a broad range of $\bar{Q}$ values of this order of magnitude, we approximately find $\mu\left(\bar{Q}\right) \simeq 101\cdots102\,\textrm{GeV}$.
As a consequence, Eq.~\eqref{eq:matching} turns into the following constraint on the parameter space of the type-I seesaw model,
\begin{align}
\pTr{yy^\dagger}\,M_1^2 \sim \left(1300\,\textrm{GeV}\right)^2 \,.
\end{align}
Once this relation is satisfied, the type-I seesaw induces a Higgs mass term of the right magnitude and with the correct sign in the Lagrangian of the classically scale-invariant Standard Model.


\section{Analysis and discussion}
\label{sec:results}


\subsection{Analytical estimates}
\label{subsec:estimates}


In Sec.~\ref{sec:seesaw}, we have separately discussed the generation of SM neutrino masses, the baryon asymmetry of the Universe, and the EW scale.
In the present section, we will now put together the pieces of the puzzle and identify the viable parameter region where all three features of the type-I seesaw can be realized at the same time.
To this end, we will make use of our results in Eqs.~\eqref{eq:CIP}, \eqref{eq:etaB0final}, and \eqref{eq:matching},
\begin{align}
\label{eq:conditions}
y_{I\alpha} = \frac{\ii}{v/\sqrt{2}}\,M_I^{1/2} R_{Ii}^{\phantom{\rm T}}\, m_i^{1/2} (U^\dagger)_{i\alpha}^{\vphantom{*}} \,,\quad  \sum_{I,\alpha}\frac{C\,\varepsilon_{I\alpha}}{z_\alpha K_\alpha^{\rm eff}} = \eta_B^{\rm obs} \,,\quad \frac{\pTr{yy^\dagger}}{16\pi^2}\,M_1^2 = \mu_0^2 \,.
\end{align}
That is, we will employ the CIP for the RHN Yukawa couplings to ensure that our analysis is in agreement with the low-energy neutrino data, and we will simultaneously impose the conditions that (i) leptogenesis results in the correct value of the baryon asymmetry and that (ii) the RHN threshold corrections induce the correct Higgs mass parameter at the RHN decoupling scale.
It is interesting to note that, in the context of the 2RHN seesaw model, these conditions completely remove the parametric freedom at low energies.
Recall that, in the case of the 2RHN seesaw, the low-energy EFT contains only one complex DOF, $z = z_{\rm R} + \ii\,z_{\rm I}$, that is not accessible in experiments [see Eq.~\eqref{eq:CIP}].
For a given RHN mass spectrum, this parameter can thus be eliminated by the two conditions $\eta_B^0 = \eta_B^{\rm obs}$ and $\Delta\mu_0^2 = \mu_0^2$.
In the following, we will turn this argument around and solve the conditions in Eq.~\eqref{eq:conditions} for the values of $M_1$ and $M_2$ that are required by leptogenesis and the neutrino option as functions of $z_{\rm R}$ and $z_{\rm I}$.
In this analysis, $z_{\rm R}$ and $z_{\rm I}$ can then be regarded as the coordinates of the unconstrained theory space of all possible UV flavor models (see Sec.~\ref{subsec:neutrinos}).


We begin by examining the constraint in Eq.~\eqref{eq:matching}, which can also be written as follows,
\begin{align}
\label{eq:mm0}
\widetilde{m} = \frac{\textrm{Tr}\big[m_{\rm D}^{\vphantom{\dagger}}m_{\rm D}^\dagger\big]}{M_1} = \frac{16\pi^2}{M_1^3}\frac{v^2}{2}\,\mu_0^2 = \widetilde{m}_0 \,.
\end{align}
Here, we introduced the mass parameter $\widetilde{m}$, which is defined in a similar way as the well-known effective neutrino mass parameters $\widetilde{m}_I = \big(m_{\rm D}^{\vphantom{\dagger}}\,m_{\rm D}^\dagger\big)_{II}/M_I$~\cite{Plumacher:1996kc,Buchmuller:2004nz}.
In this formulation, the neutrino-option constraint now states that the mass parameter $\widetilde{m}$ must obtain a particular value $\widetilde{m}_0$,
\begin{align}
\widetilde{m}_0 \simeq 48\,\textrm{meV}\left(\frac{10^7\,\textrm{GeV}}{M_1}\right)^3 \bigg(\frac{\mu_0}{100\,\textrm{GeV}}\bigg)^2 \,.
\end{align}
The mass $\widetilde{m}$ can be related to the total mass in the SM neutrino sector, $m_{\rm tot} = \sum_i m_i$, as follows,
\begin{equation}
\label{eq:mtildemtot}
\frac{\widetilde{m}}{m_{\rm tot}} = \cosh\left(2z_{\rm I}\right) + \frac{\delta M}{2}\left(\cosh\left(2z_{\rm I}\right) + \delta m \cos\left(2z_{\rm R}\right)\right) \,,
\end{equation}
where we introduced the dimensionless ratios $\delta M$ and $\delta m$ to parametrize the relative mass splittings among the heavy and light neutrino mass eigenstates, respectively,
\begin{equation}
\label{eq:dMdm}
\delta M = \frac{\Delta M}{M_1} = \frac{M_2-M_1}{M_1} \,, \quad \delta m = \frac{\Delta m}{m_{\rm tot}} = \begin{cases}\left(m_3-m_2\right)/m_{\rm tot} \overset{\rm BFP}{\simeq} 7.1\times10^{-1} & \textrm{(NH)} \\ \left(m_2-m_1\right)/m_{\rm tot} \overset{\rm BFP}{\simeq} 7.5 \times 10^{-3} & \textrm{(IH)}\end{cases} \,.
\end{equation}
Here, the numerical values for $\delta m$ correspond to the NH and IH \textit{best-fit points} (BFPs) in Tab.~\ref{tab:data}.


An important implication of Eq.~\eqref{eq:mtildemtot} is that $\widetilde{m}$ turns out to be bounded from below by $m_{\rm tot}$,
\begin{align}
\widetilde{m} > m_{\rm tot} \,,
\end{align}
which is reminiscent of the inequality $\widetilde{m}_I > m_{\rm min}$, where $m_{\rm min}$ is the smallest nonzero SM neutrino mass eigenvalue (see the Appendix of Ref.~\cite{Fujii:2002jw}).%
\footnote{Note that $\widetilde{m}_1 + \widetilde{m}_2 = m_{\rm tot}\cosh\left(2z_{\rm I}\right)$, such that $\widetilde{m} \rightarrow \widetilde{m}_1 + \widetilde{m}_2$ for $\delta M \rightarrow 0$.}
Together with the neutrino-option constraint $\widetilde{m} = \widetilde{m}_0$ in Eq.~\eqref{eq:mm0}, this lower bound on $\widetilde{m}$ results in an upper bound on the RHN mass $M_1$, 
\begin{equation}
\label{eq:M1z}
M_1 = \left(\frac{8\pi^2v^2\mu_0^2}{m_{\rm tot}}\right)^{1/3} \left[\cosh\left(2z_{\rm I}\right) + \frac{\delta M}{2}\left(\cosh\left(2z_{\rm I}\right) + \delta m \cos\left(2z_{\rm R}\right)\right)\right]^{-1/3} < \left(\frac{8\pi^2v^2\mu_0^2}{m_{\rm tot}}\right)^{1/3} \,.
\end{equation}
For both NH and IH, we thus find that $M_1$ cannot obtain values larger than $10^7\,\textrm{GeV}$,
\begin{align}
\label{eq:M1max}
m_{\rm tot} \overset{\rm BFP}{\simeq}\begin{cases} 59\,\textrm{meV} & \textrm{(NH)} \\ 99\,\textrm{meV} & \textrm{(IH)}\end{cases} \qquad\Rightarrow\qquad M_1 \overset{\rm BFP}{\lesssim}\begin{cases} 9.4 \times 10^6\,\textrm{GeV} & \textrm{(NH)} \\ 7.9 \times 10^6\,\textrm{GeV} & \textrm{(IH)}\end{cases} \,.
\end{align}
$M_1$ values below this upper bound can always be realized at the cost of a larger value of $\left|z_{\rm I}\right|$.
In the limit of a small RHN mass splitting, $\delta M \ll 1$, we find the following simple relation,
\begin{equation}
\label{eq:M1NO}
\delta M \ll 1 \qquad\Rightarrow\qquad
M_1 \approx \left(\frac{8\pi^2v^2\mu_0^2}{m_{\rm tot}\cosh\left(2z_{\rm I}\right)}\right)^{1/3} \,.
\end{equation}
Therefore, restricting the range of allowed $z_{\rm I}$ values to $z_{\rm I} \in \left[-2,2\right]$ (see Sec.~\ref{subsec:neutrinos}), we recognize that a nonzero value of $\left|z_{\rm I}\right|$ can lower $M_1$ by roughly a factor of $\sqrt[3]{\cosh 4} \simeq 3$ compared to the upper bound in Eq.~\eqref{eq:M1max}. 
This fixes the range of viable $M_1$ values that are compatible with the successful generation of the Higgs mass parameter without fine-tuned cancellations in the RHN Yukawa matrix,
\begin{equation}
\label{eq:M1range}
\delta M \ll 1 \,, \quad z_{\rm I} \in \left[-2,2\right] \qquad\Rightarrow\qquad 10^{6.5}\,\textrm{GeV} \lesssim M_1 \lesssim 10^{7.0}\,\textrm{GeV} \,.
\end{equation}
Hence, all viable $M_1$ values are of $\mathcal{O}\left(10^7\right)\,\textrm{GeV}$ and spread across only half an order of magnitude.


Next, let us turn to the baryon asymmetry $\eta_B^0$ in Eq.~\eqref{eq:etaB0final}.
In our analytical discussion, we shall restrict ourselves to the resonantly enhanced part of the $CP$ asymmetry parameter $\varepsilon_{I\alpha}$ only.
That is, we will neglect the subdominant contribution $\varepsilon_{I\alpha}^{(v)}$ in Eq.~\eqref{eq:epsilonV} for simplicity and only focus on the dominant contribution $\varepsilon_{I\alpha}^{(s)}$ in Eq.~\eqref{eq:epsilonS}.
In our numerical analysis in Sec.~\ref{subsec:scan}, we will not make any such simplification, but work with the full expression for $\varepsilon_{I\alpha}$ instead.
Furthermore, it is convenient to distinguish between two different regimes regarding the $CP$ asymmetry parameter $\varepsilon_{I\alpha}^{(s)}$, depending on whether the regulators $R_{IJ}^{\rm osc}$ and $R_{IJ}^{\rm mix}$ in Eq.~\eqref{eq:RIJoscmix} are numerically relevant or not.
It is easy to see that the transition between these two regimes is controlled by the size of the RHN mass splitting, $\Delta M = M_2 - M_1$, in relation to the RHN decay widths $\Gamma_I$.
For $\Delta M \gg \Gamma_I$, the function $f_{IJ}$ in Eq.~\eqref{eq:epsilonS} does, in fact, not need to be regularized, whereas for $\Delta M \ll \Gamma_I$, regularization is crucial,
\begin{equation}
\label{eq:fIJ}
f_{IJ} \approx \textrm{sgn}\left(M_I - M_J\right)\times\begin{cases}\Gamma_J/\left(2\,\Delta M\right) & ; \quad \Gamma_{1,2} \ll \Delta M \ll M_{1,2} \\ 2\,\Gamma_J\,M_I\,M_J\,\Delta M/R_{IJ} & ; \quad \Delta M \ll \Gamma_{1,2} \ll M_{1,2}\end{cases} \,.
\end{equation}
This behavior of $f_{IJ}$ determines the dependence of the final baryon asymmetry $\eta_B^0$ on the mass splitting $\Delta M$. 
For $\Delta M \gg \Gamma_I$ and $\Delta M \ll \Gamma_I$, the baryon asymmetry will scale as $\eta_B^0 \propto 1 / \Delta M$ and $\eta_B^0 \propto \Delta M$, respectively.
Let us now discuss these two cases in more detail one by one.


In the first case, $\Gamma_I \ll \Delta M$, we can expand $\varepsilon_{I\alpha}^{(s)}$ up to linear order in the RHN decay rates $\Gamma_I$,
\begin{align}
\label{eq:epsilonRHS}
\Gamma_I \ll \Delta M \qquad\Rightarrow\qquad
\sum_I\varepsilon_{I\alpha}^{(s)} \approx A_\alpha\,\frac{M_1\Gamma_2 + M_2\Gamma_1}{M_2^2-M_1^2}\,,\quad A_\alpha = \frac{4\,\textrm{Re}\left[\left(yy^\dagger\right)_{12}\right]\textrm{Im}\big[y_{1\alpha}^*y_{2\alpha}^{\vphantom{*}}\big]}{\left(yy^\dagger\right)_{11}\left(yy^\dagger\right)_{22}} \,,
\end{align}
where we dropped all higher-order terms in $\Gamma_I$ as well as all terms that are not resonantly enhanced.
Making use of the CIP, the Yukawa prefactor $A_\alpha$ in this expression can be written as follows,
\begin{align}
A_\alpha = \frac{4\,\delta m\,\sin\left(2z_{\rm R}\right)\left(B_\alpha\,\sinh\left(2z_{\rm I}\right) - C_\alpha\,\zeta\cosh\left(2z_{\rm I}\right)\right)}{\delta m^2\cos^2\left(2z_{\rm R}\right)-\cosh^2\left(2z_{\rm I}\right)} \,,
\end{align}
where $B_\alpha$ and $C_\alpha$ capture the dependence on the PMNS matrix $U$ and the SM neutrino masses $m_i$,
\begin{equation}
B_\alpha = \sum_i\left|U_{\alpha i}\right|^2 \frac{m_i}{m_{\rm tot}} \,,\qquad C_\alpha = \frac{2\sqrt{m_k m_l}}{m_{\rm tot}}\,\textrm{Im}\left[U_{\alpha k}^{\vphantom{*}}U_{\alpha l}^*\right] \,, \quad \left(k,l\right) = \begin{cases}\left(2,3\right) & \textrm{(NH)} \\ \left(1,2\right) & \textrm{(IH)} \end{cases} \,.
\end{equation}
These quantities also allow us to rewrite the effective washout parameter $K_\alpha^{\rm eff}$ as a function of the CIP parameter $z$.
In doing so, we can simply approximate $M_2 \approx M_1$, since $K_\alpha^{\rm eff}$ does not experience any resonant enhancement in the limit of a small RHN mass splitting.
We thus obtain
\begin{equation}
\label{eq:Kaeffapprox}
M_2 \approx M_1 \qquad\Rightarrow\qquad
K_\alpha^{\rm eff} \approx \kappa_\alpha\,\frac{m_{\rm tot}}{\zeta\left(3\right)m_*}\left(B_\alpha\,\cosh\left(2z_{\rm I}\right)-C_\alpha\,\zeta\sinh\left(2z_{\rm I}\right)\right) \,.
\end{equation}
Note that this expression is also valid in the $\Delta M \ll \Gamma_I$ regime.
$m_*$ in Eq.~\eqref{eq:Kaeffapprox} is a benchmark value for the SM neutrino masses that is sometimes referred to as the equilibrium mass~\cite{Buchmuller:2004nz}.
It allows one to express the RHN decay parameter $K_I$ in terms of the mass ratio $\widetilde{m}_I/m_*$, such that $\widetilde{m}_I \gg m_*$ and $\widetilde{m}_I \ll m_*$ are synonymous to the strong-washout and weak-washout scenarios, respectively, 
\begin{equation}
K_I = \frac{\widetilde{m}_I}{\zeta\left(3\right)m_*} \,, \quad m_* = \left(\frac{\pi^2\,g_{*,\rho}}{90}\right)^{1/2} \frac{4\pi v^2}{M_{\rm Pl}} \simeq 1.1\,\textrm{meV} \,.
\end{equation}
Combining Eqs.~\eqref{eq:epsilonRHS} and \eqref{eq:Kaeffapprox}, we arrive at the following estimate for the final baryon asymmetry,
\begin{equation}
\label{eq:etaB0comb}
\eta_B^0 \approx C\,\frac{M_1\Gamma_2 + M_2\Gamma_1}{M_2^2-M_1^2}\frac{\zeta\left(3\right)m_*}{m_{\rm tot}}\frac{4\,\delta m\,\sin\left(2z_{\rm R}\right)}{\delta m^2\cos^2\left(2z_{\rm R}\right)-\cosh^2\left(2z_{\rm I}\right)} \sum_{\alpha}\frac{D_\alpha}{z_\alpha \kappa_\alpha} \,,
\end{equation}
where the dependence on the PMNS matrix and the SM neutrino masses is now encoded in $D_\alpha$,
\begin{equation}
\label{eq:Dalpha}
D_\alpha = \frac{B_\alpha\,\sinh\left(2z_{\rm I}\right) - C_\alpha\,\zeta\cosh\left(2z_{\rm I}\right)}{B_\alpha\,\cosh\left(2z_{\rm I}\right)-C_\alpha\,\zeta\sinh\left(2z_{\rm I}\right)} \,.
\end{equation}
This estimate is dominated by the leading $1/\delta M$ term when expanding in powers of $\delta M$,
\begin{equation}
\label{eq:etaB0MMz}
\delta M \ll 1 \qquad\Rightarrow\qquad \eta_B^0 \approx \frac{C}{\delta M}\frac{\zeta\left(3\right)m_*M_1}{2\pi v^2}\frac{\delta m\,\sin\left(2z_{\rm R}\right)\cosh\left(2z_{\rm I}\right)}{\delta m^2\cos^2\left(2z_{\rm R}\right)-\cosh^2\left(2z_{\rm I}\right)} \sum_{\alpha}\frac{D_\alpha}{z_\alpha \kappa_\alpha} \,,
\end{equation}
which is our final result for $\eta_B^0$ as a function of $\delta M$, $M_1$, $z_{\rm R}$, $z_{\rm I}$, etc.\ in the $\Gamma_I \ll \Delta M$ regime.


As expected, $\eta_B^0$ scales like one inverse power of the RHN mass splitting, $\eta_B^0 \propto 1/ \delta M$.
On the other hand, it is linearly proportional to the mass splitting in the SM neutrino spectrum, $\eta_B^0 \propto \delta m$, to leading order in $\delta m$.
This is similar to the situation in standard hierarchical leptogenesis, where the DI bound on the total $CP$ asymmetry parameter also turns out to be proportional to the SM neutrino mass splitting~\cite{Davidson:2002qv}. 
An immediate consequence of this proportionality, $\eta_B^0 \propto \delta m / \delta M$, is that, given the numerical values of $\delta m$ in Eq.~\eqref{eq:dMdm}, an inverted SM neutrino mass ordering always requires a RHN mass splitting that is roughly two orders of magnitude smaller than in the case of a normal SM neutrino mass ordering, $\delta M_{\rm IH} \sim 10^{-2}\,\delta M_{\rm NH}$.
Therefore, if we speculate that UV flavor models may be biased towards larger RHN mass splittings, we are able to conclude that resonant leptogenesis tends to indicate a preference for the NH scenario rather than the IH scenario.
This is an interesting result in light of the current low-energy neutrino data, which results in a $\chi^2$ difference between the NH and IH global fits of $\Delta \chi^2 = \chi_{\rm IH}^2 - \chi_{\rm NH}^2 \simeq 9.3$~\cite{Esteban:2018azc,nufit}.
Together, these observations motivate us to assign higher priority to the NH case in the following than to the IH case.


In our numerical parameter scan in Sec.~\ref{subsec:scan}, we will be interested in assessing the maximal RHN mass splitting that is compatible with the observed value of the baryon asymmetry.
For this reason, let us now consider the special case $z_{\rm R} = \pi/4$, which maximizes $\left|\eta_B^0\right|$ as a function of $z_{\rm R}$,
\begin{equation}
\label{eq:etaB0zR}
\eta_B^0 \approx -\frac{C\,\delta m}{\cosh\left(2z_{\rm I}\right)\delta M}\frac{\zeta\left(3\right)m_*M_1}{2\pi v^2} \sum_{\alpha}\frac{D_\alpha}{z_\alpha \kappa_\alpha} \,.
\end{equation}
From this expression, it is evident that $\left|\eta_B^0\right|$ decreases exponentially as a function of $\left|z_{\rm I}\right|$ for $\left|z_{\rm I}\right| \gg 1$,
\begin{align}
\label{eq:BAUzI}
\left|z_{\rm I}\right| \gg 1 \quad\Rightarrow\qquad \eta_B^0 \propto e^{-2\left|z_{\rm I}\right|} \,.
\end{align}
This suppression can be traced back to the suppression of the $CP$ asymmetry parameter in the case of large (and fine-tuned) Yukawa couplings, $\varepsilon_{I\alpha} \propto 1/\left(yy^\dagger\right)_{II} \propto e^{-2\left|z_{\rm I}\right|}$ for $\left|z_{\rm I}\right| \gg 1$ [see Eq.~\eqref{eq:zI}].
The simple dependence of $\eta_B^0$ on $\cosh\left(2z_{\rm I}\right)$ in Eq.~\eqref{eq:etaB0zR} allows us to finally combine our analysis of resonant leptogenesis with our previous discussion of the neutrino option.
According to Eq.~\eqref{eq:M1NO}, the neutrino option implies an approximate one-to-one relation between $M_1$ and $\cosh\left(2z_{\rm I}\right)$, such that 
\begin{equation}
\label{eq:etaB0box}
\boxed{\quad\eta_B^0 \approx -\frac{C}{16\pi^3}\frac{\delta m}{\delta M}\frac{\zeta\left(3\right)m_*\,m_{\rm tot}\,M_1^4}{v^4\mu_0^2} \sum_{\alpha}\frac{D_\alpha}{z_\alpha \kappa_\alpha} \,,\quad \Gamma_{1,2} \ll \Delta M \ll M_{1,2} \,.\quad}
\end{equation}
This estimate for the baryon asymmetry is one of the main results in this paper.
It illustrates in one compact expression how the type-I seesaw model succeeds in unifying the physics of neutrino masses, leptogenesis, and EWSB.
Roughly speaking, one may summarize the content of Eq.~\eqref{eq:etaB0box} as follows:
The factors $\delta m$, $m_{\rm tot}$, and $D_\alpha$ represent the masses and flavor oscillations of the light SM neutrinos at low energies; the factors $C$, $\delta M$, $\zeta\left(3\right)m_*$, $M_1$, $z_\alpha$, and $\kappa_\alpha$ account for resonant leptogenesis in the early Universe; and the factors $v$ and $\mu_0$ reflect the spontaneous breaking of EW symmetry at the EW scale as well as the RHN threshold corrections to the Higgs mass parameter.


Based on Eq.~\eqref{eq:etaB0box}, we are now able to estimate the RHN mass splitting $\delta M$ that is required for successful baryogenesis.
For illustration, let us set the RHN mass $M_1$ to its maximally allowed value [see Eq.~\eqref{eq:M1max}] and maximize $\delta M$ over the $CP$-violating phases $\delta$ and $\sigma$, while keeping the PMNS mixing angles fixed at their respective best-fit values given in Tab.~\ref{tab:data}.
We thus find
\begin{equation}
\label{eq:deltaM}
M_1 = \begin{cases} 9.4 \times 10^6\,\textrm{GeV} & \textrm{(NH)} \\ 7.9 \times 10^6\,\textrm{GeV} & \textrm{(IH)}\end{cases} \qquad\Rightarrow\qquad \delta M \overset{\rm BFP}{\simeq}\begin{cases} 2.1 \times 10^{-4} & \textrm{(NH)} \\ 1.3\times 10^{-6} & \textrm{(IH)}\end{cases} \,.
\end{equation}
which translates into absolute mass splittings of $\Delta M \simeq 2.0\times10^3\,\textrm{GeV}$ (NH) and $\Delta M \simeq 10\,\textrm{GeV}$ (IH), respectively.
Note that these values mark the upper boundaries of the viable $M_1$ and $\delta M$ ranges, meaning in particular that the largest possible RHN mass splitting that we find is of $\mathcal{O}\left(10^3\right)\,\textrm{GeV}$.
If we allow the RHN mass $M_1$ to vary by half an order of magnitude [see Eq.~\eqref{eq:M1range}], the strong dependence of $\eta_B^0$ on $M_1$ in Eq.~\eqref{eq:etaB0box}, $\eta_B^0 \propto M_1^4$, implies a variation of $\delta M$ by two orders of magnitude.
The NH case is therefore characterized by a relative RHN mass splitting in the range $\delta M \sim 10^{-6}\cdots 10^{-4}$, while in the IH case, we expect a mass splitting in the range $\delta M \sim 10^{-8}\cdots 10^{-6}$.


Next, let us consider the $\Delta M \ll \Gamma_I$ regime, where the shape of the function $f_{IJ}$ in the $CP$ asymmetry parameter $\varepsilon_{I\alpha}^{\rm (s)}$ is determined by the regulators $R_{IJ}^{\rm osc}$ and $R_{IJ}^{\rm mix}$ [see Eqs.~\eqref{eq:epsilonS} and \eqref{eq:RIJoscmix}].
We discuss this regime mostly for completeness. 
From a conceptual point of view, it is less appealing than the $\Gamma_I \ll \Delta M$ regime because of the strong dependence on the regularization procedure, which introduces a certain degree of theoretical uncertainty; and from a phenomenological point of view, it is less appealing because of the tiny RHN mass splitting that is required for successful baryogenesis.
The only difference in the case of an extremely small mass splitting is that one should not expand $\varepsilon_{I\alpha}^{\rm (s)}$ in powers of $\Gamma_I$ [see Eq.~\eqref{eq:epsilonRHS}], but rather in powers of $\delta M$.
To leading order, this results in
\begin{equation}
\Delta M \ll \Gamma_I \qquad\Rightarrow\qquad
\sum_I\varepsilon_{I\alpha}^{(s)} \approx A_\alpha\,\delta M\,\frac{8\pi v^2}{m_{\rm tot}\,M_1} \left(\frac{X}{\cosh^2\left(2z_{\rm I}\right) - \delta m^2} + \frac{1}{X}\right) \,,
\end{equation}
where $X$ is a shorthand notation for the following function of $z_{\rm R}$ and $z_{\rm I}$ in the complex $z$ plane,
\begin{equation}
X = \frac{\cosh^2\left(2z_{\rm I}\right)-\delta m^2\cos^2\left(2z_{\rm R}\right)}{2\cosh\left(2z_{\rm I}\right)} \,.
\end{equation}
Apart from this, all steps in the calculation remain the same.
In analogy to Eq.~\eqref{eq:etaB0MMz}, we thus obtain
\begin{equation}
\eta_B^0 \approx C\,\delta M\,\frac{32\pi\,v^2\zeta\left(3\right)m_*}{m_{\rm tot}^2 M_1} \left(\frac{X}{\cosh^2\left(2z_{\rm I}\right) - \delta m^2} + \frac{1}{X}\right)\frac{\delta m\,\sin\left(2z_{\rm R}\right)}{\delta m^2\cos^2\left(2z_{\rm R}\right)-\cosh^2\left(2z_{\rm I}\right)} \sum_{\alpha}\frac{D_\alpha}{z_\alpha \kappa_\alpha} \,.
\end{equation}
In order to facilitate the comparison with Eq.~\eqref{eq:etaB0zR}, let us evaluate this expression at $z_{\rm R} = \pi / 4$,
\begin{equation}
\eta_B^0 \approx -\frac{C\,Y\,\delta m\,\delta M}{\cosh^3\left(2z_{\rm I}\right)}\frac{16\pi\,v^2\zeta\left(3\right)m_*}{m_{\rm tot}^2 M_1} \sum_{\alpha}\frac{D_\alpha}{z_\alpha \kappa_\alpha} \,,\quad Y = 5 + \frac{\delta m^2}{\cosh^2\left(2z_{\rm I}\right) - \delta m^2} \,.
\end{equation}
Here, we introduced the function $Y$, which varies in the range $Y \simeq 5\cdots6$ in the NH scenario, and which is roughly constant, $Y \simeq 5$, in the IH scenario.
Finally, let us impose the neutrino-option condition, which allows us to replace $\cosh^3\left(2z_{\rm I}\right)$ by an expression in terms of $M_1$ [see Eq.~\eqref{eq:M1NO}],
\begin{equation}
\label{eq:etaB0box2}
\boxed{\quad\eta_B^0 \approx -\frac{C\,Y}{32\pi^5}\,\delta m\,\delta M\,\frac{\zeta\left(3\right)m_*\,m_{\rm tot}\,M_1^8}{v^4\mu_0^6} \sum_{\alpha}\frac{D_\alpha}{z_\alpha \kappa_\alpha} \,,\quad \Delta M \ll \Gamma_{1,2} \ll M_{1,2} \,.\quad}
\end{equation}


\begin{figure}[t]
\begin{center}
\includegraphics[width=0.88\textwidth]{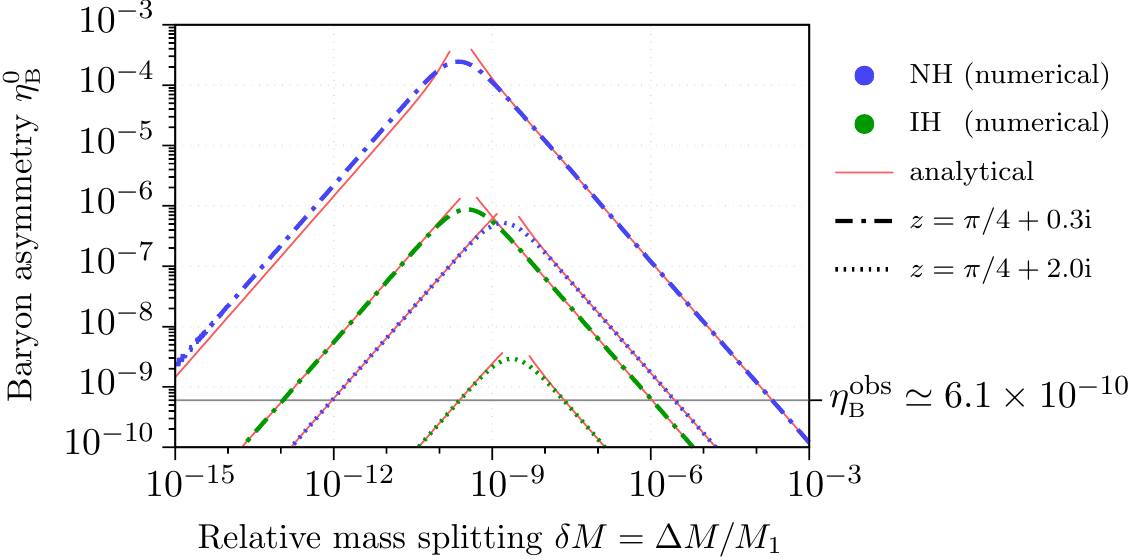}
\caption{Final baryon asymmetry $\eta_B^0$ as a function of the relative RHN neutrino mass splitting $\delta M = \Delta M / M_1 = \left(M_2-M_1\right)/M_1$ at four representative points in the complex $z$ plane.
The RHN mass $M_1$ is always chosen so as to satisfy the neutrino-option constraint in Eq.~\eqref{eq:matching}.
All other parameters are fixed at their best-fit values according to Tab.~\ref{tab:data}.
The analytical estimates at small and large $\delta M$ values are based on Eq.~\eqref{eq:etaB0box} and Eq.~\eqref{eq:etaB0box2}, respectively.
We find excellent agreement between the full numerical result and our analytical estimates, except for $\delta M$ values close to the resonance peak, where the mass splitting is of the order of the RHN decay widths, $\Delta M \sim \Gamma_{1,2}$.
Also, note how varying the CIP parameter $z_{\rm I}$ affects the solutions of the leptogenesis condition $\eta_B^0 = \eta_B^{\rm obs}$.}
\label{fig:bau}
\end{center}
\end{figure}


The expression in Eq.~\eqref{eq:etaB0box2} is the $\Delta M \ll \Gamma_I$ equivalent of our $\Gamma_I \ll \Delta M$ result in Eq.~\eqref{eq:etaB0box}.
The main difference between these two results is the different scaling with $\delta M$, $M_1$, and $\mu_0$.
While the result in Eq.~\eqref{eq:etaB0box} scales as $\eta_B^0 \propto \delta M^{-1}\mu_0^{-2}M_1^4$, we now find $\eta_B^0 \propto \delta M\mu_0^{-6}M_1^8$.
For $\Delta M \ll \Gamma_I$, the baryon asymmetry is thus enhanced by a factor $M_1^4/\mu_0^4$ and suppressed by a factor $\delta M^2$ compared to the expression in the $\Gamma_I \ll \Delta M$ regime.
This leads to a situation where successful baryogenesis can only be achieved for a RHN mass splitting that is significantly smaller than anything that we have encountered thus far, $\delta M \ll 10^{-8}$.
In our numerical parameter scan in Sec.~\ref{subsec:scan}, we will therefore ignore the alternative solution in Eq.~\eqref{eq:etaB0box2} and focus on our first solution in Eq.~\eqref{eq:etaB0box} instead.
In Fig.~\ref{fig:bau}, we show a comparison of our analytical estimates in Eqs.~\eqref{eq:etaB0box} and \eqref{eq:etaB0box2} with the full numerical result for the baryon asymmetry based on Eqs.~\eqref{eq:etaB0final}.
We find that, within their respective ranges of applicability, both of our analytical expressions are in excellent agreement with the exact result.


\subsection{Numerical parameter scan}
\label{subsec:scan}


Let us now cross-check and extend our analytical results by means of a full-fledged numerical analysis.
As in the previous section, we are going to consider the conditions in Eq.~\eqref{eq:conditions}; this time, however, we will refrain from applying any simplifying approximations and work with the full expressions that we derived in Sec.~\ref{sec:seesaw} instead.
In a first step, we wish to perform a scan of our model in the complex $z$ plane.
We are specifically interested in the region $z_{\rm R} \in \left[0,\pi\right)$ and $z_{\rm I}\in\left[-2,+2\right]$ (see Sec.~\ref{subsec:neutrinos}).
At each point in this region, we would like to solve the leptogenesis and neutrino-option constraints in Eq.~\eqref{eq:conditions} for the RHN mass $M_1$ and the RHN mass splitting $\delta M = \Delta M / M_1 = \left(M_2-M_1\right)/M_1$ and determine the \textit{largest} possible mass splitting $\delta M_{\rm max}$ that is compatible with the current low-energy neutrino data.
The fact that we decide to follow this strategy is motivated by our theoretical prejudice that exceptionally small mass splitting may be hard to come by in concrete UV flavor models.
That is, if the size of the RHN mass splitting should, \textit{e.g.}, be related to the quality of some approximate global flavor symmetry in the ultraviolet, we would expect a bias towards larger mass splittings that do not require a strong suppression of symmetry-breaking effects at higher energies.

\begin{figure}[t]
\begin{center}
\includegraphics[width=0.88\textwidth]{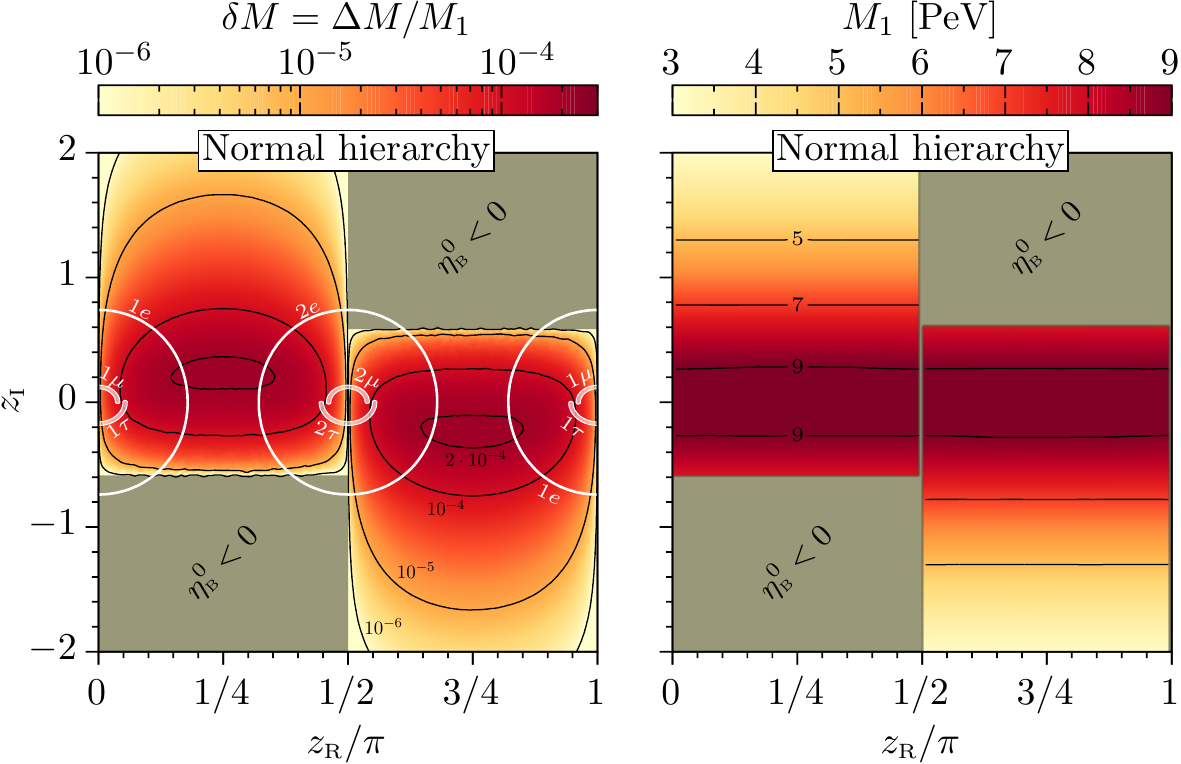}
\caption{Maximally allowed RHN mass splitting $\delta M_{\rm max}$ \textbf{(left panel)} and corresponding RHN mass $M_1$ \textbf{(right panel)} as functions of $z_{\rm R}$ and $z_{\rm I}$ in the NH scenario after varying the low-energy neutrino observables within their $3\,\sigma$ confidence intervals.
The white contour lines in the plot on the left-hand side indicate the orbits of the texture-zero flavor models discussed in Sec.~\ref{subsec:discussion}.}
\label{fig:NH}
\end{center}
\end{figure}

In order to find the maximal mass splitting $\delta M_{\rm max}$ as a function of $z_{\rm R}$ and $z_{\rm I}$, we allow the low-energy observables in Tab.~\ref{tab:data} to vary within their $3\,\sigma$ confidence ranges.
As for the $CP$-violating phase $\delta$, which is currently only poorly constrained by the data, we consider the full range of possible values, $\delta \in \left[0,2\pi\right)$. 
Similarly, we allow the $CP$-violating phase $\sigma$, which is completely unconstrained at the moment, to vary in the full range $\sigma \in \left[0,\pi\right)$.
In practice, we implement the variation of the low-energy observables by drawing random numbers from the respective ranges of allowed values. 
Likewise, we let the discrete parameter $\zeta$ in Eq.~\eqref{eq:CIP} randomly flip between $\zeta = +1$ and $\zeta = -1$.
At each point in the complex $z$ plane, we consider $10^6$ different combinations of possible values for the low-energy observables, and for each of these combinations, we solve the constraints in Eq.~\eqref{eq:conditions} for $M_1$ and $\delta M$.
In Fig.~\ref{fig:deltaM}, we present our results for $\delta M_{\rm max}$ for both the NH and IH scenarios; in Fig.~\ref{fig:NH}, we show our solutions for $\delta M_{\rm max}$ and $M_1$ in the NH case next to each other.
In view of our numerical results in Figs.~\ref{fig:deltaM} and \ref{fig:NH}, several comments are in order:


\begin{figure}[t]
\begin{center}
\includegraphics[width=0.88\textwidth]{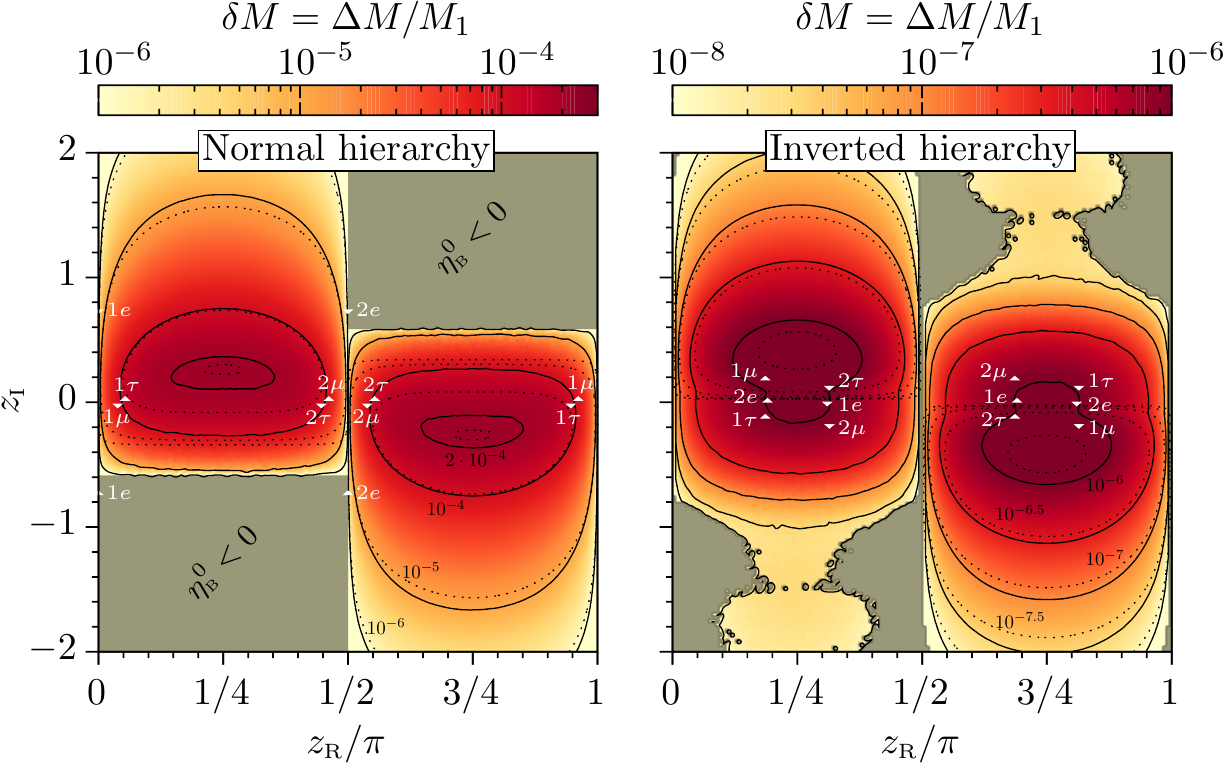}
\caption{Maximally allowed RHN mass splitting $\delta M_{\rm max}$ as a function of $z_{\rm R}$ and $z_{\rm I}$ in the NH scenario \textbf{(left panel)} and in the IH scenario \textbf{(right panel)}.
The solid contour lines represent the full numerical result after varying the low-energy neutrino observables within their $3\,\sigma$ confidence intervals, while the dotted contour lines correspond to fixed low-energy input parameters ($\delta = 3\pi/2$, $\sigma = 0$, and all other observables set to their best-fit values in Tab.~\ref{tab:data}).
The white triangles indicate the locations of the texture-zero flavor models discussed in Sec.~\ref{subsec:discussion} at the same benchmark point.}
\label{fig:deltaM}
\end{center}
\end{figure}


\begin{enumerate}
\item All plots in Figs.~\ref{fig:deltaM} and \ref{fig:NH} display the symmetry and reflection properties that we anticipated in Eq.~\eqref{eq:reflections}.
Our results are thus invariant under the different operations on the RHN Yukawa couplings shown in Eq.~\eqref{eq:reflections}.
In particular, we observe that the dependence of $\delta M_{\rm max}$ on $z_{\rm R}$ is mostly controlled by the $\sin\left(2z_{\rm R}\right)$ term in Eq.~\eqref{eq:etaB0MMz}, which stems from the fact that the $CP$ asymmetry parameter $\varepsilon_{I\alpha}^{\rm (s)}$ in Eq.~\eqref{eq:epsilonRHS} is proportional to $\textrm{Re}\left[\left(yy^\dagger\right)_{12}\right] \propto \sin\left(2z_{\rm R}\right)$.
For $2\,z_{\rm R}/\pi \in \mathbb{Z}$, it is thus impossible to realize successful baryogenesis, which explains why we fail to find solutions to the conditions in Eq.~\eqref{eq:conditions} on this hypersurface in parameter space.
\item The dependence of $\delta M_{\rm max}$ on $z_{\rm I}$ is mostly controlled by the by the $\cosh\left(2z_{\rm I}\right)$ term in Eq.~\eqref{eq:etaB0zR}, which reflects the fact that $\varepsilon_{I\alpha}^{\rm (s)}$ is inversely proportional to $\left(yy^\dagger\right)_{II} \propto \cosh\left(2z_{\rm I}\right)$.
Large values of $\left|z_{\rm I}\right|$ therefore lead to large and fine-tuned Yukawa couplings that suppress the $CP$ asymmetry parameter $\varepsilon_{I\alpha}^{\rm (s)}$.
This suppression can only be compensated for by a smaller RHN mass splitting.
On top of that, the factor $D_\alpha$ in Eq.~\eqref{eq:Dalpha} also contributes to the dependence of $\delta M_{\rm max}$ on $z_{\rm I}$,
\begin{equation}
D_\alpha = \frac{\tanh\left(2z_{\rm I}\right) - E_\alpha}{1-E_\alpha\tanh\left(2z_{\rm I}\right)} \,, \quad E_\alpha = \frac{\zeta\,C_\alpha}{B_\alpha} \,.
\end{equation}
The factor $D_\alpha$ has only little impact on the overall magnitude of the final baryon asymmetry; its main effect is that it modulates the sign of the LH-lepton-doublet asymmetry $\eta_{L_\alpha}^{\rm lptg}$.
In the region $0 < z_{\rm R} < \pi/2$, we find that $\eta_{L_\alpha}^{\rm lptg}$ obtains positive values for $z_{\rm I} > z_{\rm I}^\alpha = \textrm{artanh}\left(E_\alpha\right)/2$, while it obtains negative values for $z_{\rm I} < z_{\rm I}^\alpha$.
In the NH case, this behavior leads to a sign flip of the total baryon asymmetry at $z_{\rm I}$ values around $z_{\rm I}^{(0)} \simeq -0.6$, which is consistent with the values that one typically finds for the three critical $z_{\rm I}$ values $z_{\rm I}^e$, $z_{\rm I}^\mu$, and $z_{\rm I}^\tau$.
As a consequence, we are not able to construct viable solutions to the conditions in Eq.~\eqref{eq:conditions} for $z_{\rm I} < z_{\rm I}^{(0)}$ in the region $0 < z_{\rm R} < \pi/2$ and for $z_{\rm I} > -z_{\rm I}^{(0)}$ in the region $\pi/2 < z_{\rm R} < \pi$ in the NH case.
In the IH scenario, a similar effect occurs; however, in this case, the dependence of the excluded region on $z_{\rm R}$ and $z_{\rm I}$ is more complicated.
We will comment on this further below.
\item The $z$ dependence of $M_1$ in Fig.~\ref{fig:NH} is dictated by the relation in Eq.~\eqref{eq:M1z}.
The dependence on $z_{\rm R}$ in this relation is suppressed by the small value of $\delta M$, such that $M_1$ effectively turns out to be a function of $z_{\rm I}$ only, $M_1 \propto \cosh^{-1/3}\left(2z_{\rm I}\right)$ [see Eq.~\eqref{eq:M1NO}].
This dependence is a direct consequence of the neutrino-option constraint in Eq.~\eqref{eq:matching}, which can also be written as
\begin{equation}
\delta M\ll1 \qquad\Rightarrow\qquad 8\pi^2 v^2\mu_0^2 = \textrm{Tr}\big[m_{\rm D}^{\vphantom{\dagger}}m_{\rm D}^\dagger\big] M_1^2 \approx \cosh\left(2z_{\rm I}\right)m_{\rm tot}\,M_1^3 \,.
\end{equation}
Here, the $\cosh\left(2z_{\rm I}\right)$ factor is a consequence of the relation $\left(yy^\dagger\right)_{II} \propto \cosh\left(2z_{\rm I}\right)$, while the cubic power of the RHN mass follows from the two powers of $M_1$ in Eq.~\eqref{eq:matching} and the single power of $M_1$ that relates $\textrm{Tr}\big[m_{\rm D}^{\vphantom{\dagger}}m_{\rm D}^\dagger\big]$ to the effective mass parameter $\widetilde{m}$ [see Eqs.~\eqref{eq:mm0} and \eqref{eq:mtildemtot}].


\begin{figure}[t]
\begin{center}
\includegraphics[width=0.88\textwidth]{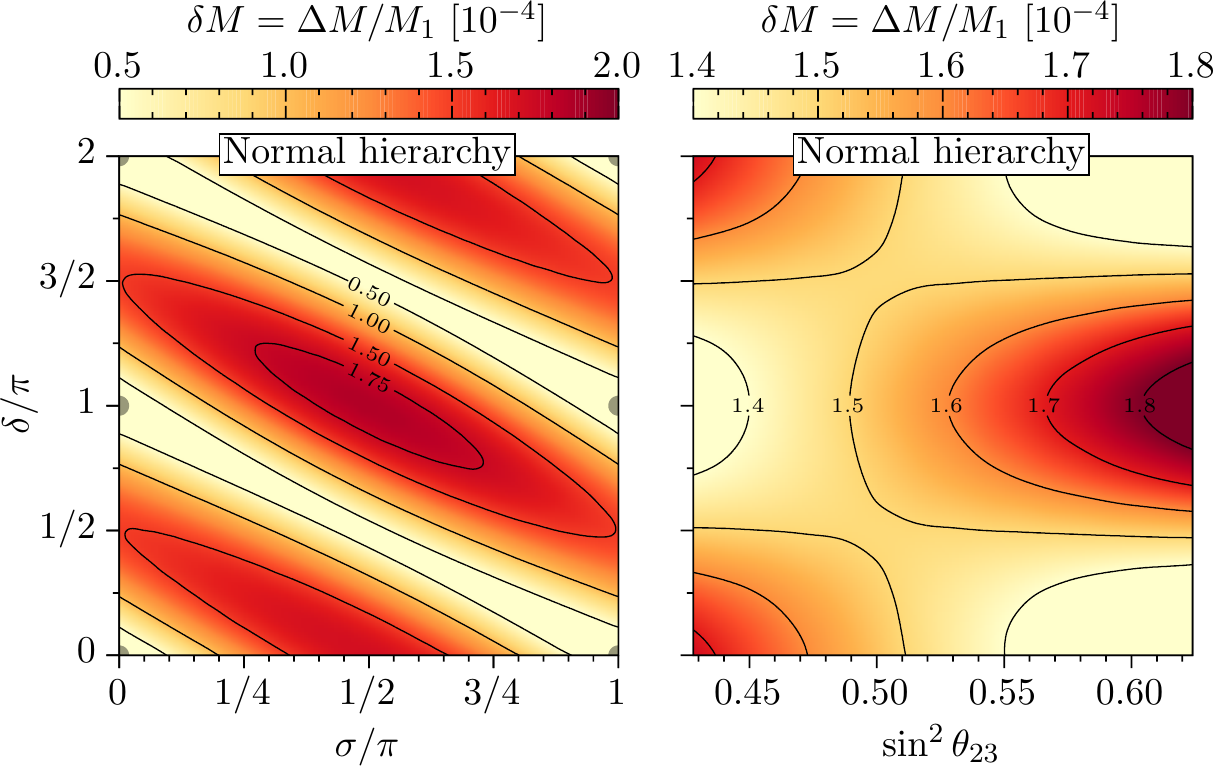}
\caption{Maximally allowed RHN mass splitting $\delta M_{\rm max}$ as a function of $\sigma$ and $\delta$ \textbf{(left panel)} and of $\sin^2\theta_{23}$ and $\delta$ \textbf{(right panel)} for $z_{\rm I} = 0$, \textit{i.e.}, for vanishing $CP$-violating phases at high energies, after varying all other low-energy neutrino observables within their $3\,\sigma$ confidence intervals.}
\label{fig:imz0}
\end{center}
\end{figure}


\item The dotted contour lines in Fig.~\ref{fig:deltaM} are the outcome of a restricted analysis based on $\delta = 3\pi/2$ and $\sigma = 0$ and all other observables kept fixed at their best-fit values in Tab.~\ref{tab:data}.
While these contour lines are also generated by our numerical code, we are able to confirm that they can be reproduced to excellent precision by the analytical expressions in Eqs.~\eqref{eq:M1NO} and \eqref{eq:etaB0MMz}.
In large parts of the complex $z$ plane, the dotted contour lines represent a good approximation of the solid contour lines, which correspond to our full numerical results after varying all low-energy input parameters.
This indicates that the variation of the low-energy observables within their $3\,\sigma$ confidence ranges only has a small effect on the outcome of our analysis in a large part of parameter space.
A notable exception to this statement occurs in the IH case in Fig.~\ref{fig:deltaM}.
For $0 < z_{\rm R} < \pi/2$ and $z_{\rm I} \lesssim -1 $ as well as for $\pi/2 < z_{\rm R} < \pi$ and $z_{\rm I} \gtrsim 1$, we find a set of solutions that cannot be reproduced by simply assuming fixed input values for the low-energy observables.
We explicitly checked that this deviation is not a numerical artifact, but a genuine outcome of our numerical algorithm. 
We thus conclude that, in this region of parameter space, the marginalization over the low-energy observables is essential.
\item Overall, we find very good agreement between the results in Figs.~\ref{fig:deltaM} and \ref{fig:NH} and our analytical estimates in Sec.~\ref{subsec:estimates}.
As expected, the mass splitting varies in the range $\delta M_{\rm max} \sim 10^{-6}\cdots 10^{-4}$ in the NH scenario and in the range $\delta M_{\rm max} \sim 10^{-8}\cdots 10^{-6}$ in the IH scenario.
Because of this relative suppression by a factor $10^{-2}$ (and because of the slight experimental preference for a normal SM neutrino mass ordering), we will no longer consider the IH case from now.
Regarding our numerical results for $M_1$, we find perfect agreement with Eqs.~\eqref{eq:M1max} and \eqref{eq:M1NO} by construction.
All $M_1$ values are of $\mathcal{O}\left(10^7\right)\,\textrm{GeV}$ and spread across half an order of magnitude.
\item As mentioned earlier, the CIP parameters $z_{\rm R}$ and $z_{\rm I}$ can be regarded as the coordinates of the unconstrained theory space of flavor models that might act as UV completions of the 2RHN seesaw.
To illustrate this statement by means of a simple example, we show the position of certain (toy) flavor models in Figs.~\ref{fig:deltaM} and \ref{fig:NH}.
We will elaborate on this in more detail in Sec.~\ref{subsec:discussion}.
\end{enumerate}


Finally, we conclude this section by pointing out that our scenario also manages to account for the generation of the baryon asymmetry even if the $CP$-violating phases $\delta$ and $\sigma$ in the PMNS matrix are the only source of $CP$ violation during leptogenesis.
This observation allows us to conclude that our scenario is compatible with the concept of leptogenesis from low-energy $CP$ violation~\cite{Branco:2001pq,Frampton:2002qc,Endoh:2002wm,Branco:2002xf,Pascoli:2006ie,Branco:2006ce,Pascoli:2006ci,Hagedorn:2017wjy,Shimizu:2017vwi,Dolan:2018qpy,Moffat:2018smo}.
To show that this is the case all we have to do is to fix the CIP parameter $z_{\rm I}$ at $z_{\rm I} = 0$ in our numerical analysis.
This renders the matrix $R$ in Eq.~\eqref{eq:CIP} real, such that $\delta$ and $\sigma$ are the only nontrivial complex phases that enter the RHN Yukawa couplings $y_{I\alpha}$.%
\footnote{The overall factor $\ii$ in Eq.~\eqref{eq:CIP} corresponds to a global phase that does not affect any of our results.}
Once $z_{\rm I}$ is set to $z_{\rm I} = 0$, we are able to perform a similar analysis as before.
We scan the complex $z$ plane along the real $z_{\rm R}$ direction, solve the conditions in Eq.~\eqref{eq:conditions} for $\delta M$ and $M_1$ for random combinations of values of the low-energy observables, and determine the maximally allowed mass splitting $\delta M_{\rm max}$.


The outcome of this analysis is shown in Fig.~\ref{fig:imz0}, where we project our results for $\delta M_{\rm max}$ into the $\sigma$\,--\,$\delta$ plane as well as into the $\sin^2\theta_{23}$\,--\,$\delta$ plane.
Our motivation for picking these projections is that $\sin^2\theta_{23}$, $\delta$, and $\sigma$ represent important observables that are expected to become better constrained by experiments in the near future~\cite{Tanabashi:2018oca}. 
In both plots in Fig.~\ref{fig:imz0}, $\delta M_{\rm max}$ varies only slightly around values of $\mathcal{O}\left(10^{-4}\right)$.
This is in accord with our results in Fig.~\ref{fig:deltaM}, where we also consistently find $\delta M_{\rm max} \sim 10^{-4}$ along the real $z_{\rm R}$ axis, and demonstrates once more the comparatively small impact of varying the low-energy observables within their $3\,\sigma$ confidence ranges.
The main message from Fig.~\ref{fig:imz0} is that it is possible to solve the conditions in Eq.~\eqref{eq:conditions} for $\delta M$ and $M_1$ across the entire ranges of $\sin^2\theta_{23}$, $\delta$, and $\sigma$ values.
In particular, it is always possible to obtain the correct sign of the baryon asymmetry, independently of the values of $\delta$ and $\sigma$.
The only exception to this statement are small regions around the following six singular points, where the baryon asymmetry vanishes identically: $\left(\delta,\sigma\right) \in \left\{\left(0,0\right),\left(\pi,0\right),\left(2\pi,0\right),\left(0,\pi\right),\left(\pi,\pi\right),\left(2\pi,\pi\right)\right\}$.
At these points, $CP$ invariance is not violated, such that there is no source for the baryon asymmetry~\cite{Sakharov:1967dj}.
In their immediate vicinity, there is not enough $CP$ violation to reproduce the observed value of the baryon asymmetry.
Other than that, it is always straightforward to realize leptogenesis from low-energy $CP$ violation in our scenario.


\subsection{Flavor models and dark matter}
\label{subsec:discussion}


Our scan of the complex $z$ plane in Sec.~\ref{subsec:scan} can be understood as a scan over all possible UV flavor models that are consistent with the low-energy neutrino data.
In the following, we will illustrate this point by means of a simple example\,---\,flavor models that are characterized by a single texture zero in RHN Yukawa matrix $y_{I\alpha}$.
The concept of texture zeros in the Yukawa sector is well known from QCD, where it can be used to successfully predict the Cabbibo angle in the quark mixing matrix~\cite{Weinberg:1977hb}.
In the SM neutrino sector, texture zeros in the Majorana mass matrix $m_{\alpha\beta}$ have been extensively studied in the literature~\cite{Frampton:2002yf,Xing:2002ta,Desai:2002sz,Merle:2006du,Dev:2006qe}.
The same is true for texture zeros in the Yukawa matrix $y_{I\alpha}$; see, \textit{e.g.}, Refs.~\cite{Frampton:2002qc,Harigaya:2012bw,Zhang:2015tea,Rink:2016vvl,Rink:2016knw} for studies of two-zero Yukawa textures in the 2RHN seesaw model.
Typically, texture zeros in a fermion Yukawa matrix are assumed to be related to a flavor symmetry at high energies (see, \textit{e.g.}, Ref.~\cite{Grimus:2004hf}), which demands that certain couplings are exactly zero or significantly suppressed compared to all other entries in the Yukawa matrix.
This flavor symmetry could, \textit{e.g.}, correspond to a Froggatt--Nielsen flavor symmetry~\cite{Froggatt:1978nt} with a flavor charge assignment such that some Yukawa couplings end up being vanishingly small.


In this section, we shall simply consider one texture zero in the RHN Yukawa matrix $y_{I\alpha}$.
Two texture zeros would only be possible, if we assumed an inverted SM neutrino mass hierarchy; more than two texture zeros are always in conflict with the low-energy neutrino data in the 2RHN seesaw.
The requirement that one element in the matrix $y_{I\alpha}$ must vanish can then be used to determine the complex rotation angle $z$.
Let us denote by $z_{1\alpha}^\zeta$ and $z_{2\alpha}^\zeta$ those values of $z$ that lead to vanishing values for the couplings $y_{1\alpha}$ and $y_{2\alpha}$, respectively.
Making use of the CIP in Eq.~\eqref{eq:CIP}, we find
\begin{align}
z_{1\alpha}^\zeta = \arctan\left(-\zeta\sqrt{\frac{m_k}{m_l}}\frac{U_{\alpha k}^*}{U_{\alpha l}^*}\right) \,,\quad z_{2\alpha}^\zeta = \arctan\left(+\zeta\sqrt{\frac{m_l}{m_k}}\frac{U_{\alpha l}^*}{U_{\alpha k}^*}\right) \,,\quad \left(k,l\right) = \begin{cases}\left(2,3\right) & \textrm{(NH)} \\ \left(1,2\right) & \textrm{(IH)} \end{cases} \,.
\end{align}
These expressions depend on the PMNS matrix $U$ as well as on the SM neutrino masses $m_i$.
However, in the following, we will fix the SM neutrino masses and the PMNS mixing angles at their best-fit values in Tab.~\ref{tab:data}, such that $z_{1\alpha}^\zeta$ and $z_{2\alpha}^\zeta$ turn into functions of the $CP$-violating phases $\delta$ and $\sigma$ only.
In Fig.~\ref{fig:deltaM}, we evaluate these functions at $\delta = 3/2\pi$ and $\sigma = 0$ and indicate the locations of the twelve complex numbers $z_{1\alpha}^\zeta$ and $z_{2\alpha}^\zeta$, where $\alpha = e,\mu,\tau$ and $\zeta = \pm1$, in the complex $z$ plane.
Here, we use upwards and downwards pointing triangles to distinguish between the $\zeta = +1$ and $\zeta = -1$ solutions, respectively.
In Fig.~\ref{fig:NH}, on the other hand, we generate the full orbits of these flavor models in the complex $z$ plane by varying $\delta$ and $\sigma$ in the ranges $\delta \in \left[0,2\pi\right)$ and $\sigma \in \left[0,\pi\right)$.
In this sense, both figures demonstrate in an illustrative manner how the complex $z$ plane can be understood as a map of the landscape of possible flavor models.
Fig.~\ref{fig:deltaM} allows us, in particular, to conclude that it is not possible to simultaneously demand the following four things in our model:
(i) a successful solution to the leptogenesis and neutrino-option constraints, 
(ii) $\delta = 3/2\pi$ and $\sigma = 0$, 
(iii) a normal SM neutrino mass ordering, and 
(iv) a vanishing RHN Yukawa coupling to the LH lepton doublet $L_e$.
Meanwhile, we can read off from Fig.~\ref{fig:NH} that a vanishing Yukawa coupling $y_{Ie}$ can be easily realized as soon as we relax our assumptions on $\delta$ and $\sigma$.
These statements are just simple examples of how our plots in Sec.~\ref{subsec:scan} can help to constrain possible embeddings of the 2RHN seesaw into UV flavor models.
We expect that more complicated flavor models may give rise to a richer structure in the complex $z$ plane that would allow for even more complex conclusions.
We leave such an investigation of alternative flavor models for future work.
In particular, it would be interesting to assess whether there are entire classes of models that are preferred or ruled out by our scenario.


Finally, let us comment on dark matter.
Throughout this work, we referred several times to the possibility that, upon extending our RHN particle content by a keV-scale state, the type-I seesaw model could in addition to neutrino masses, baryon asymmetry, and EW scale also account for dark matter.
The simplest realization of keV-scale dark matter, compatible with the type-I seesaw Lagrangian employed in this work, was proposed by Dodelson and Widrow~\cite{Dodelson:1993je}.
These authors concluded that the amount of dark matter consistent with observations can be produced by collision processes, provided nonzero mixing between the active and sterile neutrino states.
However, despite its simplicity and minimality, this scenario is nowadays excluded by the combined constraints from structure formation~\cite{Schneider:2016uqi}, supernova 1987A data~\cite{Raffelt:2011nc,Arguelles:2016uwb}, and X-ray searches~\cite{Ng:2019gch}.%
\footnote{In view of the recently discovered unidentified X-ray line at around 3.55 keV \cite{Bulbul:2014sua,Boyarsky:2014jta}, X-ray searches have led to a strong interest in this type of dark matter, which could radiatively decay into pairs of photons and active neutrinos.}
The minimal scenario for keV-scale sterile-neutrino dark matter that is compatible with present limits is therefore resonant production \`a la Shi and Fuller~\cite{Shi:1998km}.
In this scenario, the mixing between active and sterile states can be enhanced because of lepton number asymmetries, which results in the well-known \textit{Mikheev--Smirnov--Wolfenstein} (MSW) effect~\cite{Wolfenstein,Mikheev:1986gs,Mikheev:1986wj}.
The viable parameter space of the Shi--Fuller mechanism is currently probed by on-going astrophysical observations (see, \textit{e.g.}, Ref.~\cite{Perez:2016tcq}).


Resonant sterile-neutrino production is a basic ingredient of the so-called \textit{neutrino minimal Standard Model} ($\nu$MSM)~\cite{Asaka:2005pn,Asaka:2005an}, which successfully combines the physics of neutrino masses, dark matter, and baryogenesis solely based on the type-I seesaw Lagrangian.
In the $\nu$MSM, the oscillations of GeV-scale sterile neutrinos are responsible for the generation of primordial lepton asymmetries via the \textit{Akhmedov--Rubakov--Smirnov} (ARS) mechanism~\cite{Akhmedov:1998qx} (see Ref.~\cite{Ghiglieri:2019kbw} for very recent work on this topic), which then set the stage for the production of keV-scale sterile-neutrino dark matter.
Given the similarity between the $\nu$MSM and our setup, it would be interesting to investigate the possibility of an extended type-I seesaw sector featuring a split RHN spectrum.
That is, if the RHN spectrum should contain states with masses in the PeV, GeV, and keV range, one could attempt to simultaneously realize (i) the generation of the EW scale via RHN threshold corrections, (ii) baryogenesis via a combination of resonant and ARS leptogenesis, and (iii) the production of keV-scale sterile-neutrino dark matter via the Shi--Fuller mechanism. 
Alternatively, one should study in more detail whether the scenario of resonant leptogenesis explored in this paper might not suffice to generate the lepton number asymmetries needed for DM production.
In this case, one would be able to bypass the ARS mechanism at lower energies and would not need to introduce GeV-scale states in the RHN spectrum.
Both of these scenarios go beyond the scope of this paper and deserve more attention in future work.
In passing, we note that keV-scale sterile-neutrino dark matter was also explored in gauge~\cite{Bezrukov:2009th,Brdar:2018sbk} and scalar extensions~\cite{Merle:2013wta,Adulpravitchai:2014xna,Arcadi:2014dca,Brdar:2017wgy,Baumholzer:2018sfb}.
The latter are particularly appealing because the realization of scale-invariant boundary conditions calls for an extended scalar sector.


\section{Conclusions and outlook}
\label{sec:conclusions}


RH neutrinos are key players in the field of BSM model building, which allow one to address several shortcomings of the Standard Model at the same time.
In this paper, we have considered two viable RHN scenarios that manage to simultaneously explain (i) the SM neutrino oscillations, (ii) the baryon asymmetry of the Universe, and (iii) the origin of all SM particle masses.
The first scenario, which we dubbed the Dirac-neutrino option, is based on the assumption that the RHN sector preserves lepton number.
In this case, neutrinos turn into massive Dirac fermions in consequence of the Higgs mechanism, baryogenesis might proceed via neutrinogenesis, and the EW scale plays the role of a universal mass scale that determines the masses of all SM particles. 
The Dirac-neutrino option, however, suffers from (i) exceptionally small RHN Yukawa couplings, (ii) a missing connection between baryogenesis at high energies and the phenomenology of neutrino oscillations at low energies, and (iii) the fact that the EW scale should likely not be considered a fundamental scale, given the absence of signals for new physics in current experiments.
Motivated by these observations, we therefore turned to a second scenario based on an alternative underlying symmetry principle.
This second scenario, which we dubbed the Majorana-neutrino option, postulates that the SM Lagrangian satisfies scale-invariant boundary conditions in the ultraviolet, whereas the RHN sector is allowed to feature Majorana masses for the RH neutrinos that explicitly break classical scale invariance and lepton number.
In this scenario, the Higgs mass parameter in the Higgs potential is forbidden at tree level and only induced via RHN one-loop threshold corrections.
As a consequence, the RHN mass scale replaces the EW scale as the input scale that translates into the masses of all SM particles. 
This observation may be taken as a sign that the scale of new physics should, in fact, not be sought close to the EW scale but rather at energies above RHN thresholds. 
The RH neutrinos should then be regarded as messengers between the Standard Model and the BSM sector in this framework.


According to the Majorana-neutrino option, the SM neutrinos obtain small masses via the type-I seesaw mechanism.
Moreover, it can be shown that the generation of the Higgs mass term (with the correct magnitude and the correct sign) requires RHN masses of $\mathcal{O}\left(10^7\right)\,\textrm{GeV}$.
Our main contribution in this paper was to show that these two features of the type-I seesaw model are compatible with baryogenesis via resonant leptogenesis.
We focused on the minimal type-I seesaw model involving only two RH neutrinos and studied resonant leptogenesis both from an analytical and a numerical perspective.
We found excellent agreement between our two approaches and concluded that resonant leptogenesis succeeds in explaining the baryon asymmetry of the Universe for an absolute RHN mass splitting as large as $\Delta M \sim 10^3\,\textrm{GeV}$, or equivalently, for a relative RHN mass splitting as large as $\delta M \sim 10^{-4}$.
These values apply in the case of a normally ordered SM neutrino mass spectrum.
In the case of an inversely ordered SM neutrino mass spectrum, we found an additional suppression by a factor of $\mathcal{O}\left(10^2\right)$.
In addition, it is interesting to note that the success of resonant leptogenesis is not contingent on the presence of additional sources of $CP$ violation at high energies.
We could show that the leptonic $CP$ violation encoded in the PMNS matrix is enough to explain the observed baryon asymmetry.
In light of these results, we arrive at the conclusion that the type-I seesaw can not only explain the masses of all known particles but also the cosmological relic density of matter.
In this sense, the type-I seesaw may be regarded as the origin of all mass and matter in the Universe.
This statement might even extend to dark matter if the type-I seesaw sector should also contain a RHN state with a mass at the keV scale whose relic density accounts for dark matter.


\newpage

The production of keV-scale sterile-neutrino dark matter in our scenario should be investigated more carefully in future work.
Similarly, it would be interesting to extend our discussion of flavor models in Sec.~\ref{sec:results} and assess which classes of flavor models turn out to be favored or disfavored by our scenario.
In this paper, we merely restricted ourselves to a class of simple texture-zero models that allowed us to illustrate the physical meaning of our parameter plots in Figs.~\ref{fig:deltaM} and \ref{fig:NH}.
In particular, it would be worthwhile to seek an embedding of the type-I seesaw in a flavor model that automatically explains the small splitting among the RHN mass eigenvalues.
Besides that, there are several further directions in which our analysis could be extended:
(i) The renormalization scheme dependence of the RHN threshold corrections should be cross-checked by explicit higher-order computations in the type-I seesaw model.
(ii) One should repeat our analysis in the 3RHN seesaw model and study to which extent this model enables one to loosen the parameter constraints that we derived in this work.
(iii) Finally, one should embed the type-I seesaw in a fully scale-invariant UV completion that explains how the RHN masses originate from the spontaneous breaking of classical scale invariance.
Such a UV-complete model will necessarily feature an extended scalar sector, and it would be interesting to study the consequences of this extended scalar sector for leptogenesis.
All of these questions are, however, beyond the scope of this work.
We conclude by emphasizing that the type-I seesaw model is a truly intriguing extension of the Standard Model.
Despite the fact that it has been around for some 40 years, it still calls for further exploration and promises further surprises.


\subsubsection*{Note added}


During the final stages of our project, we became aware of work by Ilaria~Brivio, Kristian~Moffat, Silvia~Pascoli, Serguey~T.~Petcov, and Jessica~Turner~\cite{Brivio:2019hrj} that is closely related to ours.
Here, we comment on the relation between our results and those obtained in Ref.~\cite{Brivio:2019hrj}.
The authors of~Ref.\cite{Brivio:2019hrj} arrive at the conclusion that resonant leptogenesis in the context of the Majorana-neutrino option requires a relative mass splitting of $\delta M \sim 10^{-8}$.
In addition, they derive a lower bound $M_1^{\rm min}$ on the RHN mass $M_1$ of $M_1^{\rm min} \simeq 1.2\PeV$ in the NH case as well as of $M_1^{\rm min} \simeq 2.4\PeV$ in the IH case.
According to Ref.~\cite{Brivio:2019hrj}, these values of $M_1$ are realized at $z_{\rm I}^{\rm max} \simeq 3.3$ and $z_{\rm I}^{\rm max} \simeq 2.1$, respectively, \textit{i.e.}, at values of the CIP parameter $z_{\rm I}$ larger than those that we considered in this paper.


The lower bound on $M_1$ is associated with the fact that the baryon asymmetry becomes exponentially suppressed at large values of $\left|z_{\rm I}\right|$ [see Eq.~\eqref{eq:BAUzI}].
This suppression can only be compensated for by resorting to smaller mass splitting $\delta M$ up to the point where one reaches the resonance peak in Fig.~\ref{fig:bau} [see also Eq.~\eqref{eq:fIJ}].
Beyond that point, it is no longer possible to realize successful leptogenesis because of a strongly suppressed $CP$ asymmetry parameter.
By extending our numerical parameter scan to larger values of $\left|z_{\rm I}\right|$, we are able to locate this boundary of the viable parameter region in the complex $z$ plane.
In the NH scenario, we find that $M_1$ and $\delta M$ can become as small as $M_1^{\rm min} \simeq 1.0\PeV$ and $\delta M^{\rm min} \simeq 1.0 \times 10^{-8}$ around $z_{\rm I}^{\rm max} \simeq 3.7$, while in the IH scenario, we obtain $M_1^{\rm min} \simeq 1.8\PeV$ and $\delta M^{\rm min} \simeq 5.4 \times 10^{-9}$ around $z_{\rm I}^{\rm max} \simeq 2.5$.
Our bounds on $M_1$ and $z_{\rm I}$ are thus more or less consistent with those in Ref.~\cite{Brivio:2019hrj}.
They just appear to be slightly weaker; this may be related to details such as the sphaleron conversion factor [see Eq.~\eqref{eq:Csph}] or the fact that the analysis in Ref.~\cite{Brivio:2019hrj} accounts for the running of the SM neutrino masses.
Besides that, we observe that the mass splitting quoted in Ref.~\cite{Brivio:2019hrj}, $\delta M \sim 10^{-8}$, actually corresponds to the smallest possible value that is consistent with resonant leptogenesis and the neutrino option.
As we have demonstrated in our paper, the RHN mass splitting can, in fact, become as large as $\delta M \sim 10^{-4}$ (see Fig.~\ref{fig:deltaM}).


The upper bounds on $z_{\rm I}$ can also be estimated analytically based on the expressions in Sec.~\ref{subsec:estimates}.
It is easy to show that, at large $\left|z_{\rm I}\right|$, the $CP$ asymmetry parameter $\varepsilon_{I\alpha}^{\rm (s)}$ attains a maximal resonant enhancement for $\Delta M / \Gamma_1 \simeq \Delta M / \Gamma_2 \simeq 0.616$.
In analogy to Eq.~\eqref{eq:etaB0comb} and setting $z_{\rm R} \rightarrow \pi/4$, we are therefore able to write the final baryon asymmetry at the resonance peak as follows,
\begin{equation}
\label{eq:BAUpeak}
\eta_B^0 \approx -CF\,\frac{\zeta\left(3\right)m_*}{m_{\rm tot}}\frac{2\,\delta m}{\cosh^2\left(2z_{\rm I}\right)} \sum_{\alpha}\frac{D_\alpha}{z_\alpha \kappa_\alpha} \,.
\end{equation}
Here, the numerical factor $F$ corresponds to the difference $f_{21} - f_{12}$ [see Eq.~\eqref{eq:foscmix}] that appears in the $CP$ asymmetry parameter, evaluated at the resonance peak and at large $\left|z_{\rm I}\right|$, such that $\Gamma_1 \simeq \Gamma_2$,
\begin{align}
F = \left[\frac{x}{1+x^2} + \frac{4\,x}{1 + 4\,x^2} + \mathcal{O}\left(y^2\right)\right]_{x \simeq  0.616} \simeq 1.425 \,, \quad x = \frac{\Delta M}{\Gamma_1} \,, \quad y = \frac{\Gamma_1}{M_1} \,.
\end{align}
Making use of Eq.~\eqref{eq:BAUpeak}, one can then solve the leptogenesis constraint $\eta_B^0 = \eta_B^{\rm obs}$ for $z_{\rm I}$.
For instance, for the benchmark point discussed in Sec.~\ref{sec:results} (\textit{i.e.}, for $\delta = 3\pi/2$, $\sigma = 0$, and all other observables set to their best-fit values in Tab.~\ref{tab:data}), we obtain $z_{\rm I}^{\rm max} \simeq 3.5$ and $z_{\rm I}^{\rm max} \simeq 2.3$ in the NH and IH cases, respectively.
These values agree with our numerical results up to a deviation of less than $10\,\%$, which reflects the impact of marginalizing over the low-energy input parameters in our numerical analysis.


Finally, we mention that Ref.~\cite{Brivio:2019hrj} also contains a brief discussion of the upper bound on $M_1$, which is more or less consistent with our findings.
However, regarding the required RHN mass splitting in the case of close-to-maximal $M_1$ values, the authors of Ref.~\cite{Brivio:2019hrj} refrain from performing a quantitative analysis.
Instead, they restrict themselves to the qualitative statement  that $\delta M$ does not necessarily need to be as small as $\delta M \sim 10^{-8}$ for larger $M_1$ values.
In this sense, the mass splitting stated in the abstract of Ref.~\cite{Brivio:2019hrj}, $\delta M \sim 10^{-8}$, must be understood as a lower bound on the mass splitting that is necessary for successful leptogenesis.
As we have shown in our analysis, $\delta M$ can, in fact, be as large as $\delta M \sim 10^{-4}$ and $\delta M \sim 10^{-6}$ for NH and IH, respectively.


\subsubsection*{Acknowledgements}


We are grateful to Bhupal Dev and Daniele Teresi for helpful discussions and comments on resonant leptogenesis as well as to Simon~J.~D.~King for collaboration at the early stages of this project.
We acknowledge the friendly and constructive communication with Ilaria~Brivio, Kristian~Moffat, Silvia~Pascoli, Serguey~T.~Petcov, and Jessica~Turner towards the completion of this project.
S.\,I.\ is supported by the MIUR-PRIN project 2015P5SBHT 003 ``Search for the Fundamental Laws and Constituents''.
K.\,S.\ wishes to thank the Jo\v{z}ef Stefan Institute (IJS) in Ljubljana, Slovenia for its hospitality during an extended stay in May 2019, when parts of this manuscript were written.
The stay of K.\,S.\ at IJS was supported by the Slovenian-Italian bilateral cooperation project ``The flavor of the invisible universe'', grant numbers BI-IT-18-20-002\,/\,SI18MO07.
This project has received funding\,/\,support from the European Union's Horizon 2020 research and innovation programme under the Marie Sk\l odowska-Curie grant agreement No.\ 690575 (K.\,S.).


\newpage

\bibliographystyle{utphys27mod}
\bibliography{manuscript_2}


\end{document}